\documentclass[referee,sn-mathphys,Numbered]{sn-jnl}

\usepackage{graphicx}%
\usepackage{multirow}%
\usepackage{amsmath,amssymb,amsfonts}%
\usepackage{amsthm}%
\usepackage{mathrsfs}%
\usepackage[title]{appendix}%
\usepackage{xcolor}%
\usepackage{textcomp}%
\usepackage{manyfoot}%
\usepackage{booktabs}%
\usepackage{algorithm}%
\usepackage{algorithmicx}%
\usepackage{algpseudocode}%
\usepackage{listings}%
\usepackage{subcaption}
\usepackage{hyperref}
\usepackage{listings}
\usepackage{color}

\definecolor{mygreen}{rgb}{0,0.6,0}
\definecolor{mygray}{rgb}{0.5,0.5,0.5}
\definecolor{mymauve}{rgb}{0.58,0,0.82}

\usepackage[utf8]{inputenc}
\usepackage[T1]{fontenc}

\theoremstyle{thmstyleone}%
\theoremstyle{thmstyletwo}%

\theoremstyle{thmstylethree}%

\usepackage{soulutf8}

\newcommand{\tens}[1]{\boldsymbol{#1}} 
\newcommand{\eps}{\tens{\varepsilon}}
\newcommand{\sig}{\tens{\sigma}}
\renewcommand{\d}{\mathrm{d}}

\newcommand{\trace}[1]{\mathrm{tr}\left\{#1\right\}}

\newcommand\numberthis{\addtocounter{equation}{1}\tag{\theequation}}

\begin{document}
\title{Hybrid data-driven and physics-informed regularized learning of cyclic plasticity with Neural Networks}

\author*[1]{\fnm{Stefan} \sur{Hildebrand}}\email{stefan.hildebrand@tu-berlin.de}

\author[1]{\fnm{Sandra} \sur{Klinge}}\email{sandra.klinge@tu-berlin.de}

\affil[1]{\orgdiv{Department of Structural Mechanics and Analysis}, \orgname{TU Berlin}, \orgaddress{\street{Straße des 17. Juni 135}, \city{Berlin}, \postcode{10623}, \state{Berlin}, \country{Germany}}}


\keywords{Cyclic Plasticity; Constitutive Modelling; Machine Learning; Neural Networks; NN; Physics-Informed}

\abstract{
	An extendable, efficient and explainable Machine Learning approach is proposed to represent cyclic plasticity and replace conventional material models based on the Radial Return Mapping algorithm.
	High accuracy and stability by means of a limited amount of training data is achieved by implementing physics-informed regularizations and the back stress information. The off-loading of the Neural Network is applied to the maximal extent.
	The proposed model architecture is simpler and more efficient compared to existing solutions from the literature, while representing a complete three-dimensional material model.
	The validation of the approach is carried out by means of surrogate data obtained with the Armstrong-Frederick kinematic hardening model.
	The Mean Squared Error is assumed as the loss function which stipulates several restrictions: deviatoric character of internal variables, compliance with the flow rule, the differentiation of elastic and plastic steps and the associativity of the flow rule. The latter, however, has a minor impact on the accuracy, which implies the generalizability of the model for a broad spectrum of evolution laws for internal variables. Numerical tests simulating several load cases are shown in detail and validated for accuracy and stability.
}

\maketitle

	\section{Introduction}
	Machine learning methods have already been applied in a variety of ways to describe material behavior \cite{Rosenkranz2023, dornheim2023neural, Bock2019}.
	Goal of many applications is to replace analytical constitutive models by machine learned relationships instead of a manually created model with separate fitting steps. 
	This is of special interest when highly automated material testing laboratories are available and a quick, automatized characterization of new materials shall be carried out.
	A general, extensible neural network based methodology allows for a simplified integration of various phenomena, such as hyperelasticity, plasticity, temperature influence \cite{Lei2024} or damage \cite{Miehe2010, Aydiner2024, Bartosak2024} and does not require to establish analytical models anymore.
	While artificial neural networks (NN) are able to approximate arbitrary functions \cite{Hornik1989}, in practice the computational effort, convergence and accuracy of the results highly depend on the complete setup which requires a suitable network architecture as well as sufficient, expressive and normalized training data. 
	In many cases, a good performance requires additional regularization. The physics-informed technique is typically applied in the field of computational mechanics \cite{Raissi2019, Kollmannsberger2021, hildebrand2023comparison}.
	An alternative perspective on NN is its use as a black-box model and can help to investigate influences and dependencies among the involved quantities \cite{BockKeller2021}.
	
	Conventionally, metal plasticity is described by the $J_2$ flow theory \cite{SimoHughes, KhanHuang} based on the Huber-von Mises yield condition and the Levy-Saint Venant flow rule \cite{Bland1957}. 
	Here, linear isotropic behavior is assumed in the elastic regime, whereas the notion of equivalent plastic strain is introduced to account for isotropic hardening. Moreover, the center of the von Mises yield surface projected on the stress deviator space is assumed as a function of the back stress. Some of the typical examples of corresponding evolution equations are the Voce law and the Swift law \cite{Suchocki2021} for isotropic hardening and the Frederick-Armstrong model \cite{Frederick-Armstrong2007} for kinematic hardening. Numerical implementations typically use a predictor-corrector time stepping scheme also known as Radial Return Mapping (RRM) algorithm. Within this scheme, the predictor calculates the solution for a purely elastic time step, whereas the corrector adapts it to account for the plastic evolution, if the requirement on admissible range is not fulfilled. In cases with reduced complexity, a closed form can be found for the corrector. However, for the coupling with additional phenomena such as damage, an inner iteration loop becomes necessary \cite{AyguenKlinge2021}.
	
	Since recently, Machine Learning (ML) has been investigated as alternative tool for modeling material behavior.
	The first solutions in this context deal with purely data-driven trained fully connected neural networks (FCNNs) as homogenized constitutive model for metal plasticity \cite{FurukawaHoffman2004, MohrGorji2020}. 
	On the other hand, recurrent neural networks (RNN) are employed which incorporate information from inference queries for previous time steps for the output at the current time step \cite{Kollmannsberger2021}. This is substantiated by the fact that metal plasticity deals with sequential data due to the remaining plastic deformation and hardening.
	Exemplarily, such an approach is applied for the hardening plasticity by using two separate networks to predict back stresses and drag stresses \cite{FurukawaHoffman2004}. Similarly, the plastic strain and accumulated plastic strain can be used to propagate information over time steps for modeling isotropic hardening \cite{Jang2021}. Alternatively, the strain and stress are applied as history variables \cite{HuangWriggers2020}.
	Tests with similar architectures show that this approach is prone to a lack of stability for high-dimensional NN inputs and outputs as it typically applies for three-dimensional simulations with kinematic hardening.
	
	The long short term (LSTM) \cite{Rimoli2021} or the closely related gated recurrent unit (GRU) \cite{GRU2014} are more complex architectures that can be applied to simulate history dependent material behavior. Both these architectures contain hidden state variables within the network that
	save information from the last time step.
	Accordingly, they are considered as stateful architectures. The hidden states of these NN architectures do not coincide with the hidden (internal) variables in computational plasticity, such as back stresses and back strains.
	Inspite of their advanced architecture, stateful RNNs show several disadvantages for the modeling of cyclic plasticity. Among others, they require high computational effort for training and prediction, which goes back to the sequential dependencies within the training data sets and the large solution space of the NN parameters. Additionally, stateful RNNs are not explainable without further measures.
	
	An important aspect of NN based plasticity modeling is the generation of surrogate training data as a replacement for the experimental ones. Here, typically a set of random walks in the input dimensions is conducted \cite{BonattiMohr2021}. Moreover, the application of Gaussian Random Processes guarantees their smoothness. In a combination with FEM simulations performed on a representative volume element (RVE) \cite{JoudivandSarand2024}, NN based approaches can serve as homogenization techniques \cite{Rimoli2021}.
	
	Very recently, the problems of stability and training effort are addressed in works proposing Constitutive Artificial Neural Networks (CANNs) \cite{LinkaCyron2021, Farhat2022, LinkaKuhl2023}. These architectures are developed to replace hyperelastic material models. They have the free energy as single, scalar output and the deformation gradient or another strain measure as input. After the training, the stress tensor of interest is derived by the autograd feature of NN libraries. This design a priori incorporates the thermodynamic relations for hyperelasticity and serves as a physics-informed regularization.
	The approach mentioned cannot be directly transferred for plastic behavior, as anelasticity is not uniquely described by the free energy potential. Still, the incorporation of physics-informed regularization is adopted in the present work and significantly increases the accuracy and the stability.
	
	Some other interesting approaches focus on selected quantities of plasticity, such as predicting the algorithmic consistent modulus in plane stress states \cite{ZhangMohr2020}, learning the yield locus for anisotropic materials \cite{ShoghiHartmaier2022, Shoghi2024} or prediction of the difference between a low-fidelity analytical approximation and measured residual stresses after laser shot peening \cite{BockKeller2021}.
		
	The aim of the work at hand is the development of an NN based material model replacing the evolution equations for plastic strains and the back stresses by a data-driven approach. This approach allows to account for a large variety of materials to be represented without knowing their hardening law.
	Moreover, it is suitable to incorporate further phenomena, such as anisotropic yield surfaces \cite{ShoghiHartmaier2022} and
	non-associative plastic flow. This is for example an important issue in the context of concrete damage modeling \cite{Lubliner1989, Zoric2023} where the evolution equations of conventional material models are replaced by an NN. 
	The physics-informed technique is pursued in the paper in order to avoid the instability characteristic of purely data-driven approaches.
	The plasticity belongs to history dependent processes where information on internal variables have to be passed between adjacent time steps. This can be modeled by a stateless RNN architecture when a part of the output of the NN is fed back as part of the input in the next time step.
	In the suggested architecture, recurrent quantities are controlled by the loss function to represent the plastic strain and back-stresses. This makes the network's behavior explainable and allows the NN based model to serve as a drop-in replacement for the conventional Radial Return Mapping (RRM) algorithm \cite{Suchocki2021, AyguenKlinge2021}.
	
	The paper is structured as follows. The conventional description of cyclic hardening as implemented in the RRM is presented in Sec. \ref{subsec:ConventionalHardening} and the notation for Neural Networks is outlined in Sec. \ref{subsec:NeuralNetworks}. Thereafter, the new model architecture is proposed (Sec. \ref{subsec:suggested-arch}) and the choice for regularization terms is substantiated (Sec. \ref{sec:loss-function}). Sec. \ref{subsubsec:tangent_modulus} closes the introductory part by the derivation of the algorithmic tangent modulus from the model output.
	Since the framework and objectives are defined, Sec. \ref{sec:NumExp} describes the generation of surrogate training data for dual-phase steel and the choice of hyperparameters. The results are used to compare three different approaches for regularization. The accuracy and stability of the model with the best performances are furthermore analyzed in more detail.
	The paper finishes by summarizing the findings and potential further extensions of the approach (Sec. \ref{sec:ConcOut}).
	
	\section{Methods}\label{sec:Methods}		
	
	\subsection{Mechanical modeling of cyclic hardening} \label{subsec:ConventionalHardening}
	
	The plastic behavior belongs to a group of dissipative processes with a complex mechanical model including several groups of assumptions. Different to pure elasticity where the ansatz for the free energy is sufficient to model all aspects, the theory of plasticity involves assumptions on internal variables, yield surface and evolution equations. The latter are proposed as rheological models solely based on experimental results or based on the principles of maximum dissipation in more modern approaches \cite{SimoHughes, AyguenKlinge2021}. The present paper considers the classical framework starting with the additive decomposition of strain $\eps$ into an elastic ($\eps^e$) and a plastic part ($\eps^p$) where the latter has a deviatoric character which goes back to the theory of dislocation motion \cite{Oliveira2022}
	\begin{align}
		\eps = \eps^e + \eps^p \, \, , \quad \mathrm{tr} \{\eps^p \} = 0 \quad .
	\end{align}
	The second stage of the model introduces the yield surface ($\Phi$) as a threshold between the elastic and the plastic region. The most basic model proposed by von Mises assumes the reference stress as a norm of the deviatoric part of the stress tensor and compares it to the modified yield parameter $\overline{\sigma}_Y$		
	\begin{align}
		\Phi_{VM} = ||\overline{\sig}|| - \overline{\sigma}_Y \quad , \quad \overline{\sigma}_Y = \sqrt{\frac{2}{3}} \sigma_Y \quad .
		\label{eq:1}
	\end{align}
	The stresses and elastic strains are connected by Hooke's elasticity law according to
	\begin{align}
		\sig = \mathbb{C} : \eps_e = \mathbb{C} : \left( \eps - \eps_p \right) \quad .
	\end{align}
	The assumption on the yield surface represents the core of the plasticity model and has multiple roles. Within the theory of associative plasticity it represents a basis for the evolution of plastic strains
	\begin{align}
		\d \eps^p = \d \lambda \, \frac{\partial \phi}{\partial \sig} \quad.
	\end{align}
	Here, $\d \lambda$ is the plastic multiplier calculated such that condition (\ref{eq:1}) is fulfilled for a plastic step.
	On the other hand, the RRM numerical approach uses it to distinguish the elastic and plastic regime which is achieved by implementing Karush-Kuhn-Tucker (KKT) conditions
	\begin{align}
		\d \lambda \geq 0 \,\, , \quad \phi \leq 0 \,\, , \quad \d \lambda \,\, \phi = 0 \quad .
	\end{align}
	The incorporation of hardening effects requires extensions of the basic formulation (\ref{eq:1}). In the context of isotropic hardening, this step applies a redefinition of the modified yield limit $\overline{\sigma}_Y$ as a function of equivalent plastic strain $\eps_p$
	\begin{align}
		\overline{\sigma}_Y \left( \overline{\varepsilon}^p \right) = \overline{\sigma}_{Y, 0} + R\left( \overline{\varepsilon}^p \right) 	\quad , \quad 	\overline{\varepsilon}^p = \int \d \overline{\varepsilon}^p \, , \quad \d \overline{\varepsilon}^p = \sqrt{\frac{2}{3} \d \tens{\varepsilon}^p \cdot \d \tens{\varepsilon}^p} \quad .
	\end{align}
	
	Here, the extension $R\left( \overline{\varepsilon}^p \right)$ can be defined differently as overviewed in \cite{Suchocki2021}:
	\begin{align}
		\text{Linear law}\quad& R\left( \overline{\varepsilon}^p \right) = H \overline{\varepsilon}^p \quad&& H \in \mathbb{R} &&\text{material parameter}\\ 
		\text{Voce}\quad& R\left( \overline{\varepsilon}^p \right) = Q \left( 1 - e^{-b \overline{\varepsilon}^p} \right) \quad&& Q, b \in \mathbb{R} &&\text{material parameter}\\
		\text{Swift}\quad& R\left( \overline{\varepsilon}^p \right) = C \left( \overline{\varepsilon}^p + e_0  \right)^n - k \quad&& C, e_0, n, k \in \mathbb{R} &&\text{material parameter} \,.
	\end{align}
	Note that the list of isotropic hardening models is not limited to the previous choice but a number of alternative approaches is available \cite{Suchocki2021}.
	Moreover, an extension of (\ref{eq:1}) is also possible by introducing kinematic hardening. The latter is crucial for modeling cyclic behavior and implementation of Bauschinger and ratcheting effects manifesting through the change of yield limit over the number of loading cycles \cite{AyguenKlinge2021, Reese2004}. The yield limit in this case takes the form 
	\begin{align}
		\Phi = \left|| \overline{\sig} - \tens{\chi} \right|| - \overline{\sigma}_Y(\overline{\varepsilon}^p) = 0 \, ,  \quad \tens{\chi} = \sum_{i = 1}^N \tens{\chi}_i \, \quad i \in [1, N] \, , \quad N \in \mathbb{N} \quad ,
	\end{align}
	where $\chi$ represents back stresses and is a purely deviatoric quantity ($\trace{ \tens{\chi}_i } = 0 \, $).
		
	The evolution of the back stresses is given by a hardening model, an example being the Armstrong-Frederick equations \cite{Frederick-Armstrong2007} with material parameters $C_, \gamma_i \in \mathbb{R}$
	\begin{align}
		\d \tens{\chi}_i = \frac{2}{3} C_i \, \d \eps^p - \gamma_i \, \d \overline{\varepsilon}^p \, \tens{\chi}_i \quad .
	\end{align}

	\subsection{Basics on Neural Networks} \label{subsec:NeuralNetworks}
\begin{figure}
		\centering
		\includegraphics[width=0.8\textwidth]{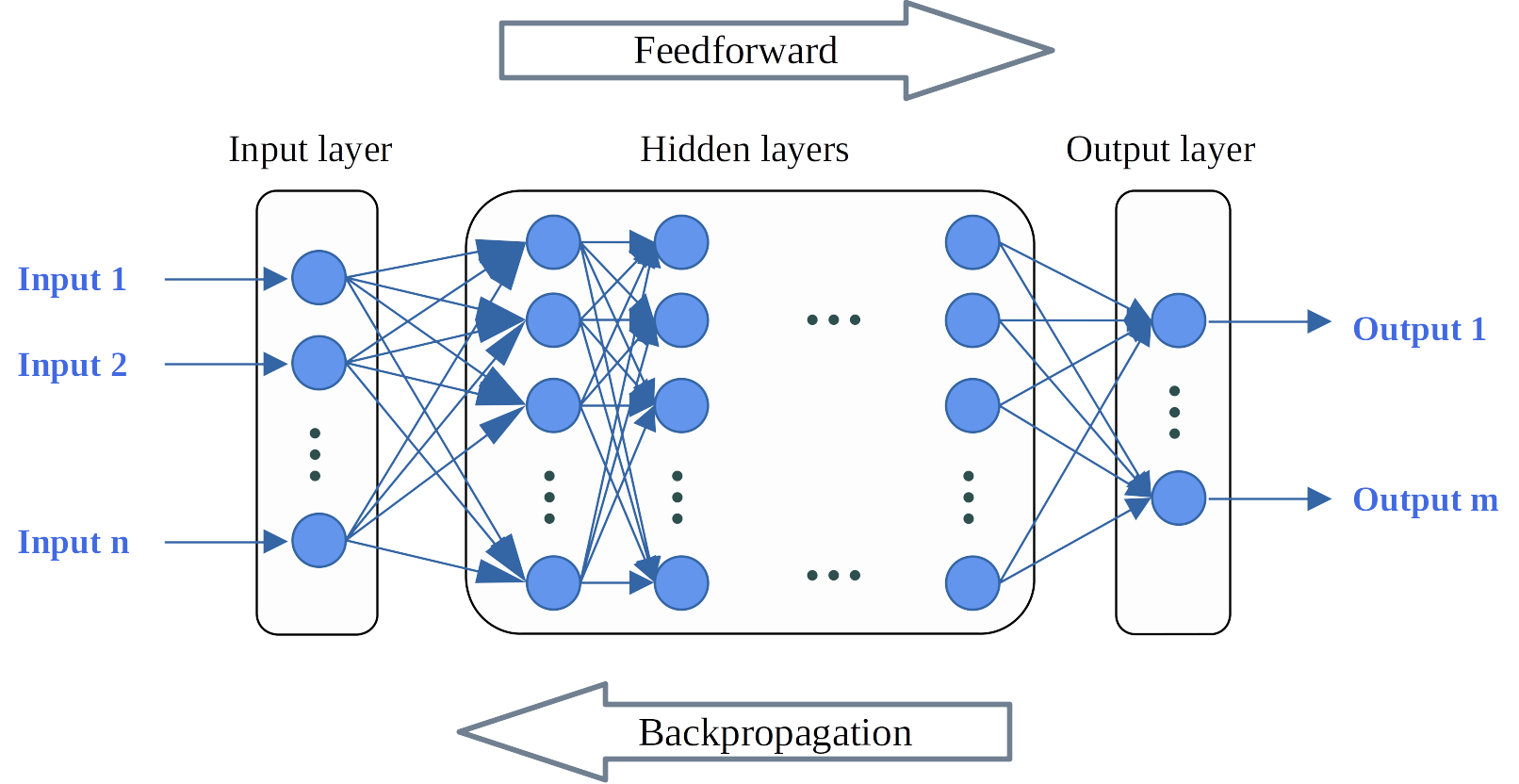}
		\caption[FCNN]{Information flow in a Fully Connected Neural Network (FCNN) (Adapted from \cite{Nguyen-Thanh2020}).}
		\label{fig:fcnn-pic}
	\end{figure}

	Artificial Neural Networks consist of layers of neurons \cite{Kollmannsberger2021} as illustrated in Fig. \ref{fig:fcnn-pic} .
	Each neuron carries out a (typically nonlinear) activation function. In case of a Fully Connected Neural Network (FCNN), each neuron receives its input as linear transformation of the outputs of the neurons in the layer before. 
	The output $\boldsymbol{\mathcal{R}}^i$ of the $i^{th}$ layer is thus calculated by
	\begin{align}
		\boldsymbol{W}^i &= \boldsymbol w^{i} \boldsymbol{\mathcal{R}}^{i-1} + \boldsymbol{b}^i \,\,, \\
		\boldsymbol{\mathcal{R}}^i &= \boldsymbol{a}^i(\boldsymbol{\Theta}^i,  \boldsymbol{W}^i) \quad .
	\end{align}	
	The weights $\boldsymbol w^{i}$ and biases $\boldsymbol{b}^i$ of the $i^{th}$ layer, together with the parameters $\boldsymbol{\Theta}^i$ of the activation functions $\boldsymbol{a}$, form the set of parameters $\theta$ of the Neural Network. Other parameters of the network architecture (like number of layers and layer widths) and the optimizer algorithm (e.g. step width) are considered hyperparameters and have to be chosen manually or by an outer optimization strategy.
	
	Layers between the input and output layer are hidden layers. The number of (hidden) layers is referred to as the network's depth, whereas the number of neurons within a layer is called the width of the layer. 
	The whole network represents an arbitrary (continuous) mapping (Universal approximation theorem, \cite{Hornik1989}) $\mathcal{R}: \mathbb{R}^n \mapsto \mathbb{R}^m$ from the input to the output side, with $n$ and $m$ the input and output layer width, respectively.
	
	To approximate a mapping $\tilde{\mathcal{R}}$ on a subset $D \subset \mathbb{R}^n$ by a network, the mapping is given indirectly by a set $T_{tr}$ of training tuples $t_{tr}^k \left(P^k, \tilde{\mathcal{R}}(P^k) \right), \,\, P^k \in D, \,\, k \in \mathbb{N}$, which together form the training data set. Here, $P^k$ are the input samples and $\tilde{\mathcal{R}}(P^k)$ the corresponding target outputs. A set $T_{te}$ of testing tuples $t_{te}^l, \,\, l \in \mathbb{N}$  is required to check the quality of the approximation the network has learned so far. Usually, $T_{tr} \cap T_{te} = \emptyset$. In the application of a network, arbitrary sets of data within $D$ can be the input, but the exact output is usually unknown and only approximated by the net.
	
	Conventionally, the parameters of the network are adapted by a gradient descent based optimization algorithm like Adam \cite{Kingma2014}. This algorithm minimizes the loss value $\mathbb{L}$ which is commonly defined as the discrepancy between the target outputs and the outputs of the network with the current parameters. A frequent choice for the loss function is Mean Square Error (MSE) for $N \in \mathbb{N}$ tuples $t^k$ 
	\begin{align}
		\mathrm{MSE} = \frac{1}{N} \sum_{k=1}^{N} \left(\tilde{\mathcal{R}}(P^k) - \mathcal{R}(P^k) \right)^2 \quad. \label{eq:MSE}
	\end{align}
	
	The optimization employs (partial) derivatives of the loss function w.r.t.\@ the parameters in the network. Machine learning networks like PyTorch therefor record all operations acting on a variable from input to output symbolically, so that fast, highly accurate derivation becomes possible. This feature is referred to as autograd \cite{Paszke2017}. Its use, however, is not limited to derivations w.r.t.\@ network parameters.
	
	By default, the NN parameters are initialized randomly before the first optimizer step.
	The following challenges need to be considered in a (purely) data-driven NN setup \cite{FurukawaHoffman2004}:
	\begin{itemize}[leftmargin=1em]
		\item The data set needs to be rich enough to
		\begin{itemize}[leftmargin=1em]
			\item span the whole intended space of input data,
			\item contain an appropriate amount of data points within the input space,
			\item include all the phenomena of interest.
		\end{itemize}
		\item The data set should not suffer from unwanted biases.
		\item The training phase might be computationally very costly.
		\item It is necessary to determine a well-suited NN architecture with enough parameters to represent the desired mapping of interest without the risk of over-fitting and divergence of the optimizer during training.
	\end{itemize}

	In case of physics-informed learning, only input quantities are created for the NN, instead of a training data set with input output tuples \cite{Krishnapriyan2021, Wang2021b, FuhgBouklas2022}. The target output is not calculated explicitly, but the NN output is fed into appropriately chosen physical constraints that are often partial differential equations (PDEs).
	The residual of the PDEs with the current NN output as estimated solutions is calculated and compiled into the loss value. This inverse form of the problem can often be faster calculated than the direct solution of the PDE. 
	In this setting, the effort for the generation of a training data set is reduced to the sampling of input data, whereas the discovery of the mapping based solely on the residuals is often significantly more complex.
	
	For the modeling of cyclic plasticity, the purely data-driven approach exhibits insufficient stability for multiple load cycles.
	On the other hand, the evolution of the hardening cannot be analytically pre-determined and usually needs to be measured. Hence, a purely physics-informed approach is not applicable, either.
	In order to address these issues, the present work suggests a hybrid approach extending the data-driven training by general assumptions about plastic flow in metals.
			
\begin{figure}[h]
		\centering
		\includegraphics[width=\textwidth]{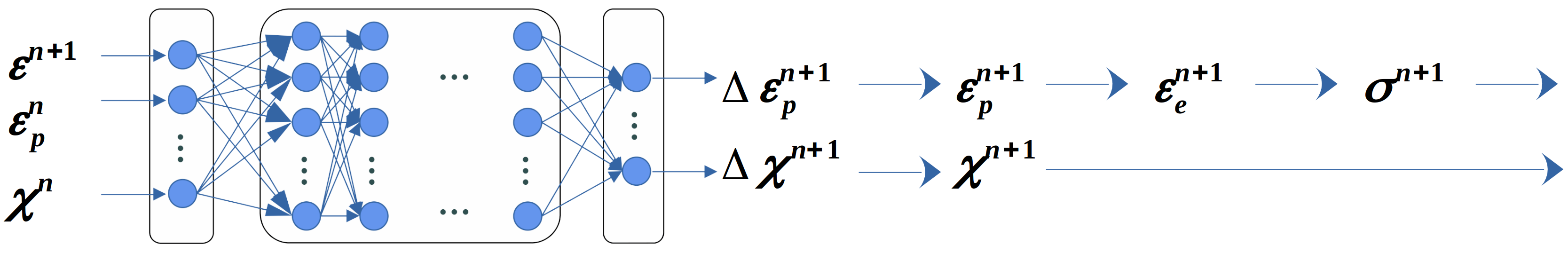}
		\caption[Schema]{Suggested NN schema.}
		\label{fig:sugg-schema}
	\end{figure}

	The method uses the current total strain $\boldsymbol{\varepsilon}^{n+1}$, the plastic strain from the last time step $\boldsymbol{\varepsilon}^{n}_p$ and the back stresses from the last time step $\tens{\chi}^{n}$ as input variables.
	Furthermore, it calculates the current values (upper index $n+1$) for plastic strains $\eps_p^{n+1}$, elastic strains $\eps_e^{n+1}$, stresses $\sig^{n+1}$ and back stresses $\tens{\chi}^{n+1}$ via the plastic strain update $\Delta \boldsymbol{\varepsilon}_p^{n+1}$ and the back stress update $\Delta \tens{\chi}^{n+1}$. The whole data flow is illustrated in Fig. \ref{fig:sugg-schema} and Tab. \ref{tab:inandoutputs}. The physics-informed regularization is introduced during training via additional terms of the loss function as described in Sec. \ref{sec:loss-function}.
	
	\begin{table}[h]
		\begin{tabular}{c|c}
			\hline
			Inputs & Targets\\
			\hline
			$\overline{\eps}_{t}$ & \\
			$\eps^p_{t-1}$ & $\Delta \eps^p$\\
			$\tens{\chi}_{t-1}$ & $\Delta \tens{\chi}$\\
			\hline
		\end{tabular}
		\caption{Inputs and outputs of the proposed NN.}
		\label{tab:inandoutputs}
	\end{table}

	\subsection{Suggested Neural Network architecture}\label{subsec:suggested-arch}
	The network is set up as a plain and stateless Fully Connected Neural Network (FCNN).
	It is provided with the current strain state as well as the plastic strain and sum of back stresses from the previous time step, which completely characterize its plastic state. The mentioned input and output quantities are similar to the ones of the conventional Radial Return Mapping algorithm which makes the network behave more explainably compared to the architectures where such quantities are implicitly learned as hidden states.
	
	An obvious choice for the output are the plastic strain $\eps_p$ and back stress $\tens{\chi}$ for the current time step. However, these quantities only show a small difference to the corresponding input quantities from the previous time step.
	The numerical tests have shown that the NN for such a constellation of input and output data only answers by repeating the input values. This drawback has motivated another choice of output quantities including the increments of plastic strains $\Delta \eps_p$ and back stresses $\Delta \tens{\chi}$.
	In this case, the target quantities are determined as a postprocessing step by
	\begin{align}
		\eps_{p,\, ML}^{n+1} &= \eps_{p,\, ML}^{n} + \Delta \eps_{p,\, ML}^{n+1} \,\,, \quad \eps_{p,\, ML}^0 = \tens{0}\\
		\overline{\eps}_{e,\, ML}^{n+1} &= \overline{\eps}^{n+1} - \eps_{p,\, ML}^{n+1} \\
		\overline{\sig}_{ML}^{n+1} &= 2 \, G \,\, \overline{\eps}_{e,\, ML}^{n+1} \,\,, \quad G = E / (2(1+\nu)) \\
		\tens{\chi}_{ML}^{n+1} &= \tens{\chi}_{ML}^{n} + \Delta \tens{\chi}_{ML}^{n+1} \quad ,
	\end{align}
 	for time step $n+1$, with the shear modulus $G$ and overlined quantities being the deviatoric parts.

	The components of an NN in PyTorch are set up for normalized input and output data by default. This applies to the sampling of initial weights as well as to the activation functions. If values with largely different orders of magnitude shall be processed, as it is the case here with strains and (back) stresses, an appropriate normalization to the same order of magnitude, preferred $10^0$, is necessary. 
	The examples carried out in this work show that the scaling of strains by $10^2$ and of back stresses by $10^{-8}$ yields good results. The difference in the order of magnitude between the scaling factors corresponds to the order of magnitude of Young's modulus of the simulated material.
	
	The accuracy of the results and the training effort can often be improved by relieving the NN as much as possible \cite{BockKeller2021}.
	In the case at hand, this is achieved considering the deviatoric strain $\overline{\eps}^{n+1}$ instead of total strains $\eps^{n+1}$. This preprocessing step relieves the NN from discovering the isochoric character of the process by itself.

	\subsection{Setup of the loss function and numerical implementation}\label{sec:loss-function}
	The suggested hybrid approach requires training data as well as regularization constraints. Both of them are included in the training via contributions to the loss function. 		
	The physics-informed regularization \cite{Raissi2019} is constructed with the requirements of
	\begin{itemize}
		\item the deviatoric character of $\eps_p$ and $\tens{\chi}$,
		\item compliance with the von Mises yield criterion,
		\item the Karush-Kuhn-Tucker conditions on the plastic multiplier and the flow rule,
		\item the associativity of the flow rule.
	\end{itemize}
	
	Similar to the conventional RRM, the trial stress $\overline{\sig}_{tr}$ is defined as the deviatoric part of the stress that would occur at the current time step if the plastic quantities from the previous time step remain unchanged.
	It is calculated from the deviatoric part of the trial elastic strain $\overline{\eps}_{e,\, tr}$.
	The implementation of the trial stress into the yield function indicates whether an update of the internal variables is necessary.
	Moreover, the trial stress is used to calculate the normal of the yield surface in the previous time step, $\tens{N}$
	\begin{align}
		\overline{\eps}_{e,\, tr} &= \overline{\eps}^{n+1} - \eps_p^n \\
		\overline{\sig}_{tr} &= 2 \, G \, \overline{\eps}_{e,\, tr}\\	
		\tens{N} &= \frac{\partial \Phi}{\partial \sig} \approx \frac{\overline{\sig}_{tr} - \tens{\chi}^{n}}{ || \overline{\sig}_{tr} - \tens{\chi}^{n} ||} \quad .
	\end{align}

	Several time steps are processed in one batch during the training process.
	To distinguish between elastic and plastic steps, two masking arrays $m_e$ and $m_p$ are introduced. The elements of $m_e$ are set to one for time steps where the yield condition is fulfilled without plastic update and are zero otherwise. The elements of $m_p$ are set to one for time steps where a plastic update is necessary and are zero otherwise. In the following, the element-wise multiplication is denoted here by '$\circ$'. Furthermore, batched quantities are not written in bold in order to distinguish them from physical tensors.

	\paragraph{Stipulating deviatoric character of internal variables}
	The plastic strain and the back stresses are deviatoric quantities which induces that their increments have to be deviatoric, too
	\begin{align}
		r\_epsp\_trace = \trace{\eps_p},\\
		r\_chi\_trace = \trace{\tens{\chi}} \quad .
	\end{align}

	\paragraph{Flow rule elastic step}
	Positive values of the yield function are not admissible
	\begin{align}
		r\_flow\_elastic = m_e \circ \max\{0 \,,\, \phi\} = m_e \circ  \max\left\{0 \,,\, \left|\left| \overline{\sig}^{ML} - \overline{\tens{\chi}}^{ML} \right|\right| - \overline{\sigma}_{Y}(\overline{\varepsilon}^p)\right\} \quad .
	\end{align}

	\paragraph{Flow rule plastic step}
	The yield function value has to be equal to zero for a plastic step. Positive values are not admissible, whereas negative values characterize elastic steps
	\begin{align}
		r\_flow\_plastic = m_p \circ \phi = m_p \circ \left( \left|\left| \overline{\sig}^{ML} - \overline{\tens{\chi}}^{ML} \right|\right| - \overline{\sigma}_Y(\overline{\varepsilon}^p) \right) \quad .
	\end{align}
	
	
	\paragraph{Increment, elastic step}
	No update of internal variables is necessary if the yield function evaluated for the trial stress is lower than or equal to zero.
	The differences are processed by the loss function separately and then summed up for all $i$ and $j$
	\begin{align}
		r\_epsp\_elastic_{ij} &= m_e \circ  \Delta \eps^{p, ML}_{ij}  \\
		r\_chi\_elastic_{ij} &=  m_e \circ \Delta \chi^{ML}_{ij} \quad .
	\end{align}
	
	\paragraph{Data driven loss}
	The network outputs are compared to the target values from the training data set. The comparison includes increments $\Delta \eps^p$ and $\Delta \tens{\chi}$ as well as the results for $\sig$ after postprocessing the network outputs
	
	\begin{align}
		d\_sig_{ij} &= \overline{\sig}^{ML}_{t, ij} - \overline{\sig}_{t, ij}  \\
		d\_depsp_{ij} &=  \Delta \eps^{p, ML}_{ij} - \Delta \eps^{p}_{ij} \\
		d\_dchi_{ij} &=  \Delta \tens{\chi}^{ML}_{ij} - \Delta \tens{\chi}^{p}_{ij} \quad .
	\end{align}

	\paragraph{Associative flow rule, plastic step (optional)}
	The model uses the associative flow rule such that the increment of plastic strain is collinear to the normal of the yield surface
	\begin{align}
		r\_assoc\_plastic &= m_p \circ \left( \frac{\eps^{ML, p}}{||\eps^{ML, p}||} - \frac{\tens{N}}{||\tens{N}||} \right) \quad .
	\end{align}

	\paragraph{Construction of the overall loss value}
	Mean Square Error (MSE) is chosen as the loss criterion since it typically yields robust and accurate results. It is applied to evaluate single loss contributions that are summed up to the overall loss. Note that MSE is applied on each tensor component individually. The loss contributions are normalized by the corresponding scaling factors of the NN input and output
	\begin{align*}
		loss =& \,\, 
		\sum_{ij} \mathrm{MSE} (d\_depsp_{ij})  + \sum_{ij} \mathrm{MSE} (d\_dchi_{ij}) + \sum_{ij} \mathrm{MSE} (d\_sig_{ij}) \\
		& +\mathrm{MSE} (r\_epsp\_trace) + \mathrm{MSE}(r\_chi\_trace) \\		
		& +\mathrm{MSE} (r\_flow\_elastic) + \mathrm{MSE} (r\_flow\_plastic) \\ 
		& +\sum_{ij} \mathrm{MSE} (r\_epsp\_elastic_{ij}) + \sum_{ij} \mathrm{MSE} (r\_chi\_elastic_{ij})  \numberthis\\
		& + \mathrm{MSE} (r\_assoc\_plastic)  \quad .
	\end{align*}

	\subsubsection{Derivation of the elastic tangent modulus as postprocessing step}\label{subsubsec:tangent_modulus}
	
	The developed model is aimed at the replacement of existing constitutive laws for plasticity and might have different applications. One of its primary purposes is the implementation in finite element codes which, among others, requires the evaluation of the elastic tangent modulus $\mathbb{E}^t$ \cite{Suchocki2021}. The latter is defined as the derivative of stresses $\sig$ w.r.t. strains $\eps$ and is an indispensable quantity for the numerical solution of solid body problems.
	By assuming Hooke's linear elasticity with 
	\begin{align}
		\sig = \mathbb{C} : \eps_e = \mathbb{C} : \left( \eps - \eps_p \right) \quad ,
	\end{align}
	where $\mathbb{C}$ is the elasticity tensor, the elastic tangent modulus turns into
	\begin{align}
		\mathbb{E}_t &= \frac{\partial \sig}{\partial \eps} = \frac{\partial \left(\mathbb{C} : \left(\eps - \eps_p\right)\right)}{\partial \eps} = \mathbb{C} : \left(\tens{1} - \frac{\partial \eps_p}{\partial \eps}\right) \quad .
	\end{align}
	The derivative $\frac{\partial \eps^p}{\partial \eps}$ can be determined automatically by the symbolic derivation function of NN packages such as autograd in PyTorch.

	\clearpage
	\section{Numerical validation}\label{sec:NumExp}
	The surrogate training and validation data are generated by using the Armstrong-Frederick model implemented in RRM as suggested in \cite{Suchocki2021}.
	The model assumes two back stress tensors and allows for isotropic hardening. The applied material parameters corresponding to dual-phase steel DP1000 and are listed in Tab. \ref{Parameter of AF-Suchocki} whereas Tab. \ref{Inout of AF-Suchocki} summarizes the input and output quantities of the model. Here, $n$ denotes the time step.
	
	\begin{table}[h]
		\begin{tabular}{c|c|l}
			\hline
			$E$ & 79308.361 MPa  & Young's modulus\\
			$\nu$ & 0.3 & Poisson's number\\
			$\sigma_{Y, 0}$ & 843.902 MPa & initial yield limit\\
			$Q_1$ & 0 & parameter for isotropic hardening\\
			$b_1$ & 0 & parameter for isotropic hardening\\
			$C_1$ & 58791.656 MPa & parameter for first kinematic hardening part\\
			$\gamma_1$ & 147.7362 & parameter for first kinematic hardening part\\
			$C_2$ & 1803.7759 MPa & parameter for second kinematic hardening part\\
			$\gamma_2$ & 0 & parameter for second kinematic hardening part\\
			$tol$ & $10^{-6}$ & tolerance for convergence of inner iteration\\
			\hline
		\end{tabular}
		\caption{Material parameters used for data generation.}
		\label{Parameter of AF-Suchocki}
	\end{table}
	\begin{table}[h]
		\begin{tabular}{c|l}
			\hline
			\textbf{Input:}  &  \\
			\hline
			$\eps^{n+1}$ & total strain tensor\\
			\hline
			& values from the previous time step\\
			$\tens{\chi}^n$ & sum of the back stress tensors\\
			$\tens{\chi}_{1}^{n}$ & first back stress part\\
			$\tens{\chi}_{2}^{n}$ & second back stress part\\
			$\eps_{p}^{n}$ & plastic part of the strain tensor\\
			$\overline{\varepsilon}_p^{n}$ & accumulated plastic strain\\
			\hline
			\textbf{Output:} & \\
			\hline
			$\tens{\chi}^{n+1}$ & updated sum of the back stress tensors\\
			$\tens{\chi}_{1}^{n+1}$ & updated first back stress part\\
			$\tens{\chi}_{2}^{n+1}$ & updated second back stress part\\
			$\eps_{p}^{n+1}$ & updated plastic part of the strain tensor\\
			$\overline{\varepsilon}_p^{n+1}$ & updated accumulated plastic strain\\
			\hline
			$\mathbb{E}_t^{n+1}$ & elastoplastic tangent modulus \\
			\hline
		\end{tabular}
		\caption{Input and output of the RRM algorithm used for data generation.}
		\label{Inout of AF-Suchocki}
	\end{table}
	
	\subsection{Hyperparameters}
	The hyperparameters are chosen by systematic variation, based on suggestions in the literature \cite{FurukawaHoffman2004, Rimoli2021, MohrGorji2020}.
	The choice of the activation functions turned out to be significantly more performant compared to common suggestions from the literature. In particular, the sigmoid function showed better performance than the tangens hyperbolicus function.

	The three succeeding optimizer settings are necessary to improve the final accuracy. 
	The Adam optimizer proved to be suitable for a robust start of the training with faster convergence towards the approximate solution, whereas the Adamax algorithm emphasizes the maximal contributions to the loss and is used for the later epochs to refine the solution with two different consecutive learning rates ($lr$) as shown in Tab. \ref{tab:hyperparameter}.
	
	Furthermore, mini-batching is employed to split the training data in several batches. Within each epoch, the network is trained once on each batch of the whole training data including the execution of the optimizer. For all data points in the batch, the gradients of the loss are summed up.
	A larger batch size allows for a faster execution and more stability of the training, whereas smaller batches allow for better accuracy. The batch size is chosen as large as possible without risking the result quality.
	The finally chosen hyperparameters are shown in Tab. \ref{tab:hyperparameter}.
	\begin{table}[h]
		\begin{tabular}{c|c}
			\hline
			Number of layers & 6\\
			\hline
			& 1 x Sigmoid\\
			Activation functions & 4 x LeakyReLU\\
			& 1 x Sigmoid\\
			\hline
			Layer width & 64\\
			\hline
			Epochs & 80\\
			\hline
			Batch size & 64\\
			\hline
			Loss function & Mean square error\\
			\hline
			& 5 epochs Adam ($lr = 0.0005$)\\
			Optimizer & 5 epochs Adamax ($lr = 0.0001$)\\
			& then: Adamax ($lr = 0.00001$)\\
			\hline
		\end{tabular}
		\caption{Hyperparameter of the Neural Network.}
		\label{tab:hyperparameter}
	\end{table}
	
	The network size is compared to the proposed LSTM architecture for the simulation of the effective behavior of heterogeneous materials with J2 plasticity \cite{Rimoli2021}. 
	The LSTM architecture uses four hidden layers with 100 neurons. In total, this makes $4 \cdot 100^2 = 40\,000$ parameters in the linear layers in addition to the parameters of the LSTM cells. The architecture in the present work only requires $6 \cdot 64^2 = 24\,576$ parameters.
	Furthermore, the LSTM architecture is trained with up to 1500 epochs, whereas the present model already shows good results after 80 epochs.

	\subsection{Generation of surrogate training data}
	The training data include stress and strain histories generated by random Gaussian processes. Six independent time series are created, one for each entry of the total strain tensor. This procedure avoids the introduction of biases in the data set and results in an equal amount of both positive and negative strain values which is necessary to account for arbitrary cyclic loading histories.
	
	The conventional RRM, which is used for the generation of surrogate training data, is only stable and accurate for sufficiently small changes in the strains between consecutive time steps. In contrast, the artificial time step size should not be too small to avoid bloating the data set with similar data points. This trade-off dictates the choice of the sampling interval.
		
	Each strain history is chosen to contain $20\,000$ data points. Since the computational effort for the data creation w.r.t. the amount of samples in each strain history increases nonlinearly, each strain history is divided into 10 patches of 2000  consecutive samples. To keep the strain history free from discontinuities, these patches are blended linearly over the first 1000 samples of the subsequent patch.
	Finally, the strain histories are scaled by the factor $10 \frac{\sigma_{Y, 0}}{E}$, which leads to a sufficient coverage of the space with plastic yielding.
	
	Comparative training runs with otherwise identic parameters show that acceptable results need at least 250 strain histories. The application of 480 strain histories yields a significant improvement in accuracy.
	For a comparison, the solution of a 2D problem proposed in \cite{Rimoli2021} requires $10\,000$ strain histories with 200 steps each.
	To summarize, the present approach requires between 250 and 480 strain histories with $20\,000$ points each for a three-dimensional problem with nonlinear hardening.
	
\begin{figure}[h]
		\centering
			\includegraphics[width=0.75\linewidth]{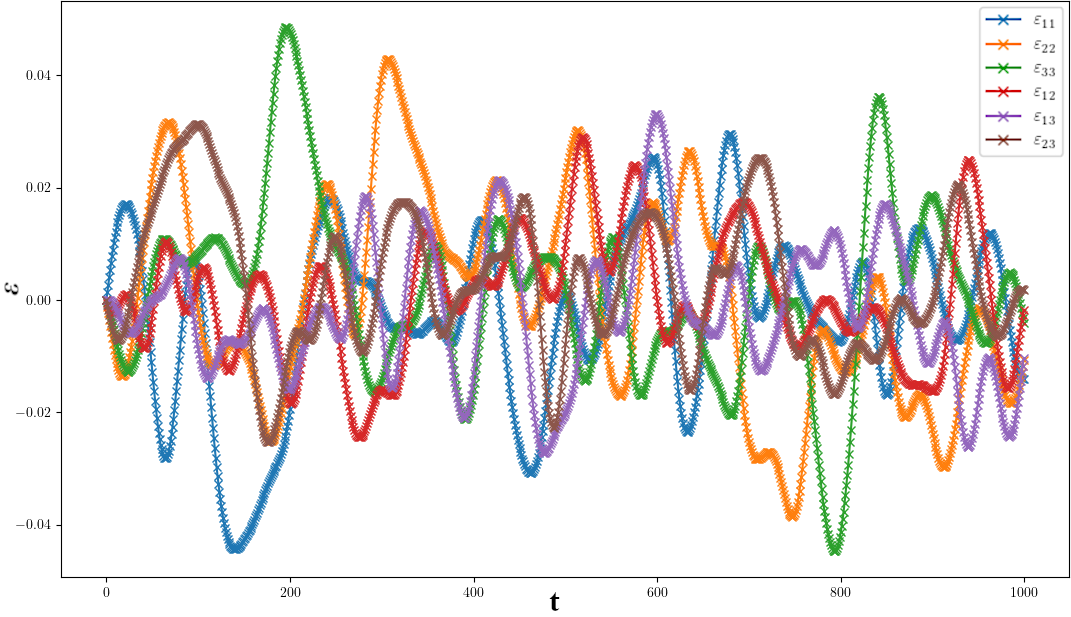}
		\caption{Components of $\eps$ used as training data. The plot is clipped on the first 1000 data points.}
		\label{fig:Traindata_time_zoomed}
	\end{figure}

	The chosen parameters for the data generation are listed in Tab. \ref{tab:datageneration}.	
	One example of the strain histories is shown in Fig. \ref{fig:Traindata_time_zoomed}.  The plot is clipped on the first 1000 data points for a better illustration.

	\begin{table}[h]
		\begin{tabular}{c|c}
			\hline
			Number random walks & 480\\
			Gaussian random process generator & Normal distribution \\
			Covariance kernel & exponentiated quadratic\\
			Mean value & 0 \\
			Sampling interval & $[-50, 50]$\\
			Samples per patch & 2000\\
			Number of patches & 10\\
			Scaling factor & $10 \frac{\sigma_{Y, 0}}{E}$\\
			\hline
		\end{tabular}
		\caption{Properties of data generation.}
		\label{tab:datageneration}
	\end{table}
	
	\subsection{Results}
	Training data are created from Random Gaussian Processes in all six independent strain components, whereas the test cases are a selection based on simple strain histories. This guarantees that the test paths are distinct from the training paths.
	Additional test cases (not shown here) with more complex loading paths and different artificial time step widths confirmed similar accuracy.
	The highest deviations are observed for pure shear as shown later on.
	
	Since the data generation is completely randomized for each strain tensor entry, the network behavior is symmetric towards the coordinate axes. Thus, only plots and loading cases in the $11$, $22$ and $12$ direction are presented from the set of all three-dimensional results.

	\paragraph{Cyclic uniaxial strain test}
	The first example considers prescribed uniaxial cyclic strains with linearly increasing amplitude
	\begin{align}
		\varepsilon_{11}(t) = f_s \left(a + \frac{t}{t_{end}}\right) \cdot \left(b - \cos(t)\right) \quad .
	\end{align}
	Here, $f_s$ is a scaling factor, $a$ is the initial proportionality constant for the amplitude and $t_{end}$ is the total time period of simulations. Displacements in all other directions except $11$ are constrained. The excitation is chosen to grow asymmetrically in tension and compression to illustrate the ratcheting effect by introducing constant $b$.
	In the simulations, the constants take the values $f_s = 5  \, \frac{\sigma_{Y}}{E}$, $a = 0.2$ and $b = 0.5$ and the time step is $\Delta t = 0.3\,\,\text{s}$.
	
	Three models (A, B and C) are trained and compared to investigate the influence of single regularization terms. Network architecture and size, training data set and hyperparameters are identical for all examples, only the loss function differs.
	
	\begin{table}
		\begin{tabular}{|r|c|c|c|}
			\hline
			Regularization & \textbf{Model A}  & \textbf{Model B} & \textbf{Model C}\\
			\hline
			Data driven loss, $\boldsymbol{\varepsilon}_p$ & \checkmark & \checkmark  &  \checkmark \\
			\hline
			Data driven loss, $\boldsymbol{\chi}$ & \checkmark & \checkmark  &  \checkmark \\
			\hline
			Data driven loss, $\boldsymbol{\sigma}$ & \checkmark & \checkmark  &  \checkmark \\
			\hline
			\hline
			Purely deviatoric $\boldsymbol{\eps}_p$ & -- & \checkmark  &  \checkmark \\
			\hline
			Purely deviatoric $\boldsymbol{\chi}$ & -- & \checkmark  & \checkmark  \\
			\hline
			Flow rule, elastic step & -- & \checkmark & \checkmark \\
			\hline
			Flow rule, plastic step & -- & \checkmark & \checkmark \\
			\hline
			No increment, elastic step & -- & \checkmark & \checkmark \\
			\hline
			Associative flow rule, plastic step & -- & -- & \checkmark \\
			\hline
		\end{tabular}
	\caption{Regularization parts included in the models for comparison.}
	\label{tab:Regularization-parts}
	\end{table}

\begin{figure}[h]
		\centering
		\begin{subfigure}[c]{0.49\linewidth}
			\centering
			\includegraphics[trim={0.5cm 23cm 41.2cm 11cm},clip,width=\linewidth]{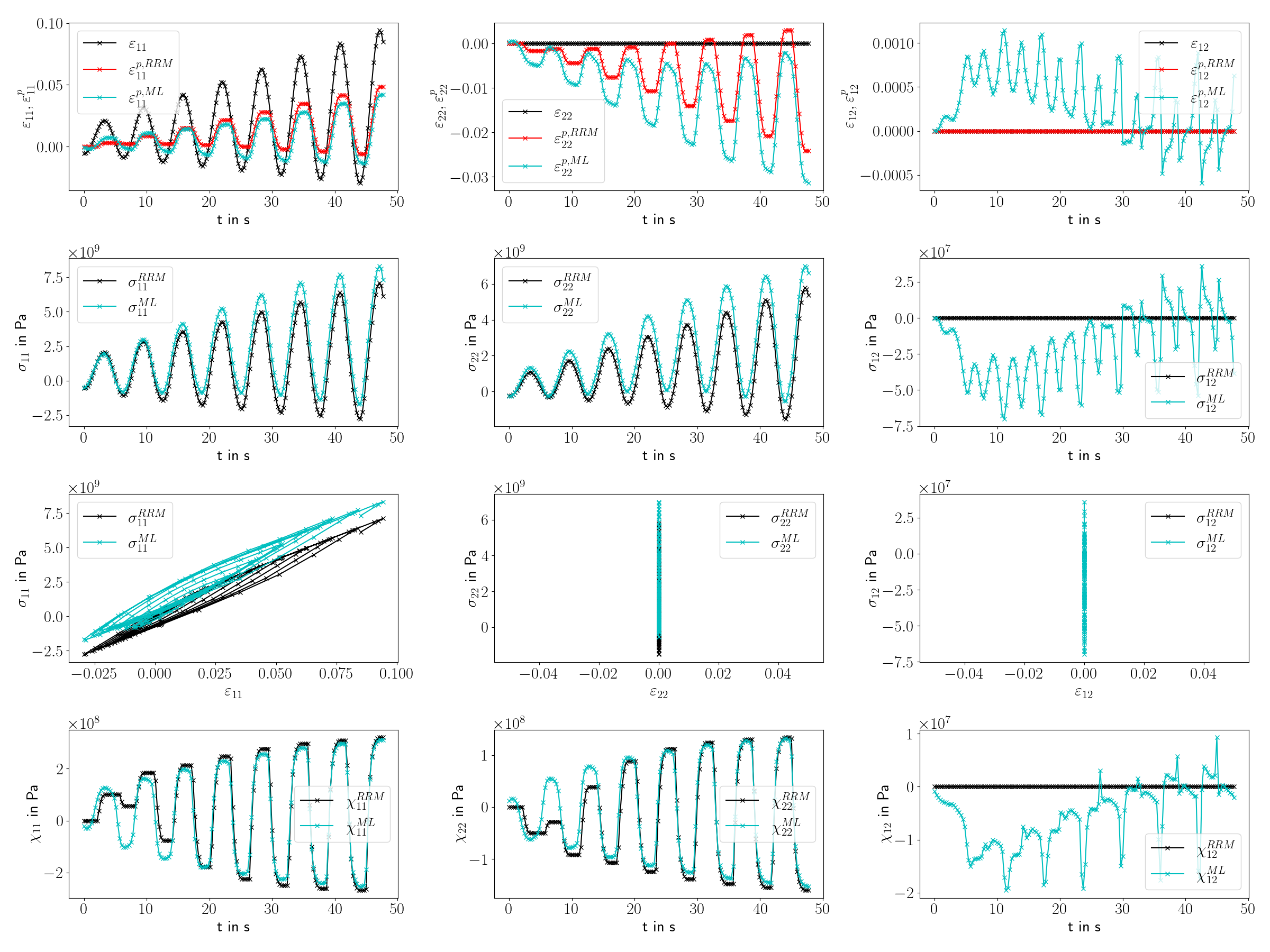}
			\subcaption{}
			\label{fig:results61002_2D}
		\end{subfigure}		
		\hfill
		\begin{subfigure}[c]{0.49\linewidth}
			\centering
			\includegraphics[trim={0.5cm 12cm 41.2cm 22.5cm},clip,width=\linewidth]{results__61002_test_No_2}
			\subcaption{}
			\label{fig:results61002_2G}
		\end{subfigure}
		
		\begin{subfigure}[c]{0.49\linewidth}
			\centering
			\includegraphics[trim={0.5cm 23cm 41.2cm 11cm},clip,width=\linewidth]{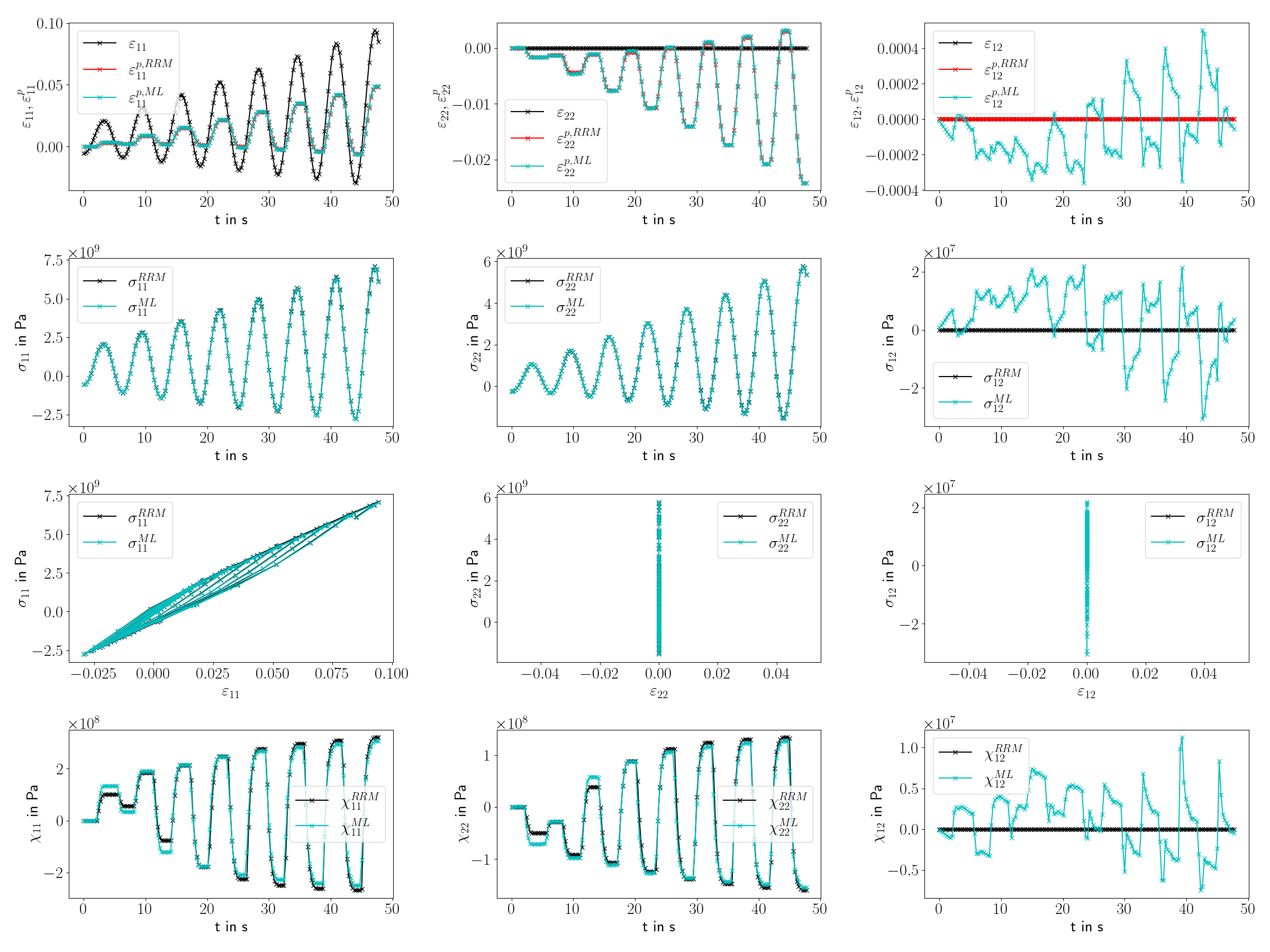}
			\subcaption{}
			\label{fig:results60102_2D}
		\end{subfigure}
		\hfill
		\begin{subfigure}[c]{0.49\linewidth}
			\centering
			\includegraphics[trim={0.5cm 12cm 41.2cm 22.5cm},clip,width=\linewidth]{results__60102_test_No_2}
			\subcaption{}
			\label{fig:results60102_2G}
		\end{subfigure}

		\begin{subfigure}[c]{0.49\linewidth}
			\centering
			\includegraphics[trim={0.5cm 23cm 41.2cm 11cm},clip,width=\linewidth]{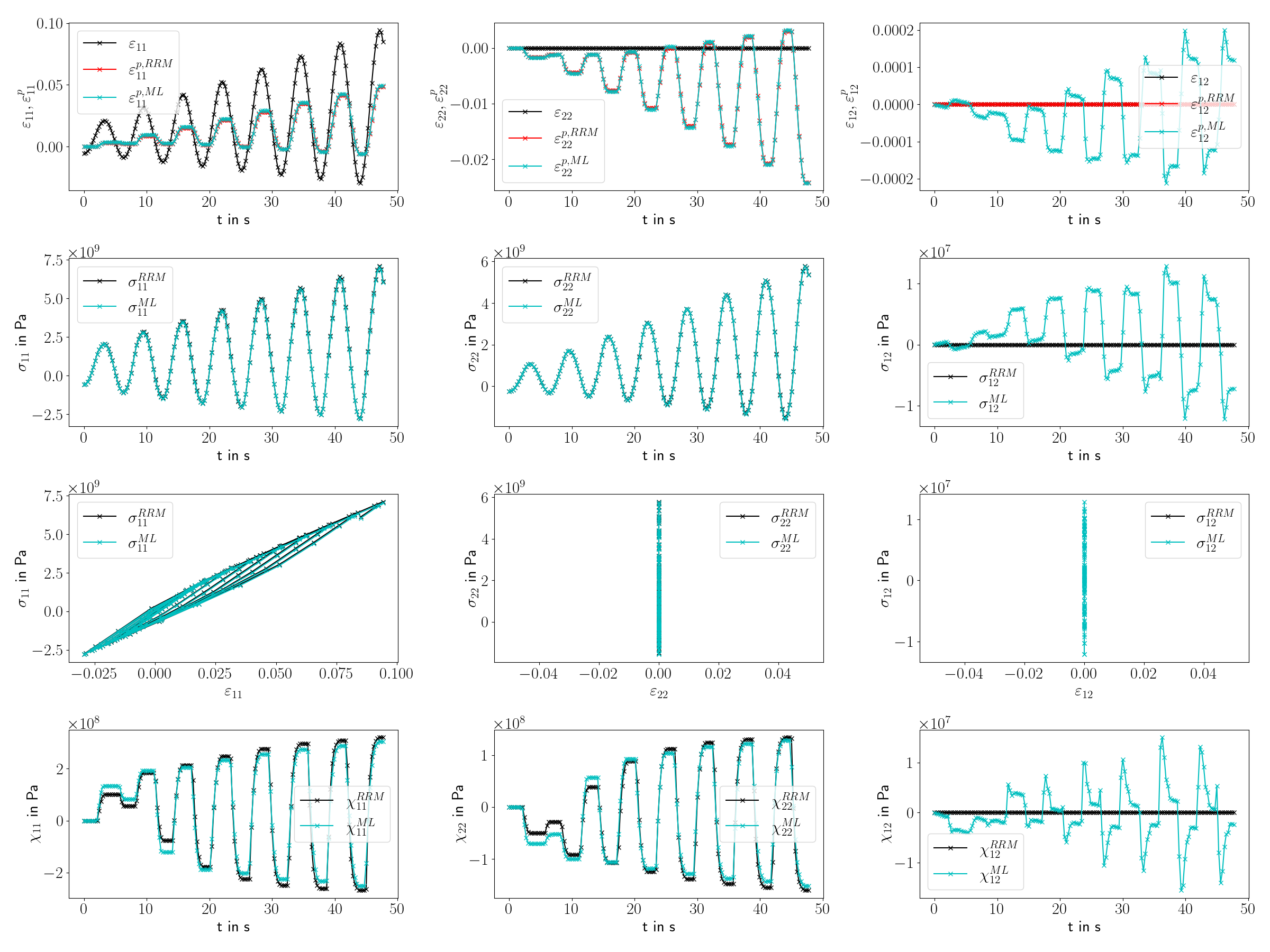}
			\subcaption{}
			\label{fig:results6002_2D}
		\end{subfigure}
		\hfill
		\begin{subfigure}[c]{0.49\linewidth}
			\centering
			\includegraphics[trim={0.5cm 12cm 41.2cm 22.5cm},clip,width=\linewidth]{results__6002_test_No_2}
			\subcaption{}
			\label{fig:results6002_2G}
		\end{subfigure}
		\caption{Comparison of stress results (ML vs. RRM) for a uniaxial deformation test in 11 direction. a) - b) Model A. c) - d) Model B. d) - e) Model C.}
		\label{fig:results_2DG}
	\end{figure}

	The results (Figs. \ref{fig:results61002_2D} and \ref{fig:results61002_2G}) show that the purely data-driven model A without additional regularizing loss terms exhibits a significant drift in the stress outputs which increases in every cycle. For that reason, the model is not applicable for the simulation of cyclic plasticity.
	In contrast, models B and C (Figs. \ref{fig:results60102_2D} - \ref{fig:results6002_2G}) achieve sufficient accuracy and high cycle stability. This is underlined by further tests with up to 50 load cycles where no signs of drift or instability are found.
	
	In a direct comparison, model B and C achieve comparable results with an equal level of accuracy. Thus, it can be conducted that a regularization with the associativity of the flow rule has a minor influence on the results and can be omitted. Moreover, it indicates that the model is suitable for material with non-associated flow.
	
	Due to the constrained lateral contraction, the stresses in $22$ direction (Fig. \ref{fig:results60102_2E}) and $33$ direction are smaller but of the same order of magnitude as the stresses in $11$ direction. Accordingly, the levels of accuracy have been found to be similar to the accuracy for the $11$ direction.

	\begin{figure}[h]
		\centering
		\begin{subfigure}[c]{0.49\linewidth}
			\centering
			\includegraphics[trim={21cm 23cm 21cm 11cm},clip,width=\linewidth]{results__60102_test_No_2}
			\subcaption{}
			\label{fig:results60102_2E}
		\end{subfigure}
		\hfill
		\begin{subfigure}[c]{0.49\linewidth}
			\centering
			\includegraphics[trim={41.5cm 23cm 0.2cm 11cm},clip,width=\linewidth]{results__60102_test_No_2}
			\subcaption{}
			\label{fig:results60102_2F}
		\end{subfigure}

		\caption{Stress results (ML vs. RRM) for a uniaxial deformation test in 11 direction, model B. a) Stress in 22 direction. b) Stress in 12 direction.}
		\label{fig:results60102_2EF}
	\end{figure}

	The shear stresses $\sig_{12}$ for the presented uni-directional deformation are expected be equal to zero as confirmed by the RRM. The ML results have oscillatory character with zero mean (Fig. \ref{fig:results60102_2F}). Compared to the normal stresses $\sig_{11}$, the outputs for this stress component are of much lower order of magnitude (less than $0.5\%$ at peak values) and can be considered as negligible.
	
\begin{figure}[h]
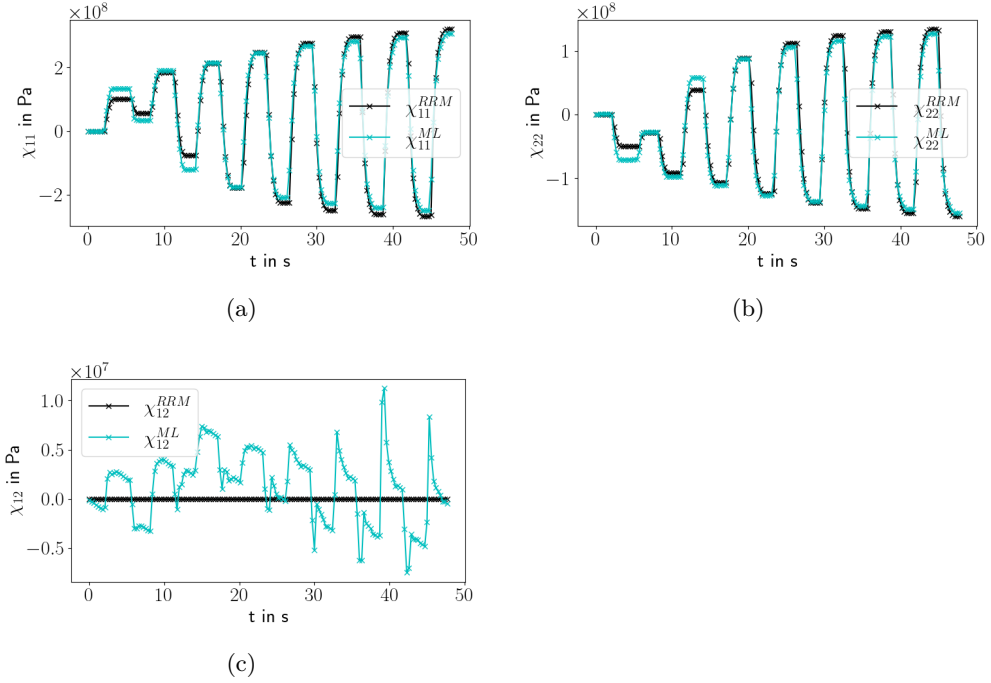

		\begin{subfigure}[c]{0.49\linewidth}
			\centering
			\includegraphics[trim={0.5cm 0.5cm 41.5cm 33.6cm},clip,width=\linewidth]{results__60102_test_No_2}
			\subcaption{}
			\label{fig:results60102_2J}
		\end{subfigure}
		\hfill
		\begin{subfigure}[c]{0.49\linewidth}
			\centering
			\includegraphics[trim={21cm 0.5cm 21cm 33.6cm},clip,width=\linewidth]{results__60102_test_No_2}
			\subcaption{}
			\label{fig:results60102_2K}
		\end{subfigure}
			\hfill
		\begin{subfigure}[c]{0.49\linewidth}
			\centering
			\includegraphics[trim={41.5cm 0.5cm 0.2cm 33.6cm},clip,width=\linewidth]{results__60102_test_No_2}
			\subcaption{}
			\label{fig:results60102_2L}
		\end{subfigure}
			
		\caption{Back stress results (ML vs. RRM) for a uniaxial deformation test in 11 direction, model B. a) Back stress in 11 direction. b) Back stress in 22 direction. c) Back stress in  12 direction.}
		\label{fig:results_2JKL}
	\end{figure}

	The stability can be evaluated by means of the back stress outputs (Fig. \ref{fig:results_2JKL}), since they serve as explainable state variables in the chosen RNN architecture. The largest deviations occur in the first cycle and reduce in the following cycles. Evaluations with up to 50 cycles with varying load amplitudes show that the deviations stabilize within 20 cycles.
	The ML model slightly underestimates back stresses $\boldsymbol{\chi}$ (deviations of $9\%$ at peak values).
	The deviations for the stresses $\boldsymbol{\sigma}$, however, are smaller and comprise typically less than $2 \%$.
	
\begin{figure}[h]
		\begin{subfigure}{0.49\linewidth}
			\centering
			\includegraphics[trim={41.5cm 23cm 0.2cm 11cm},clip,width=\linewidth]{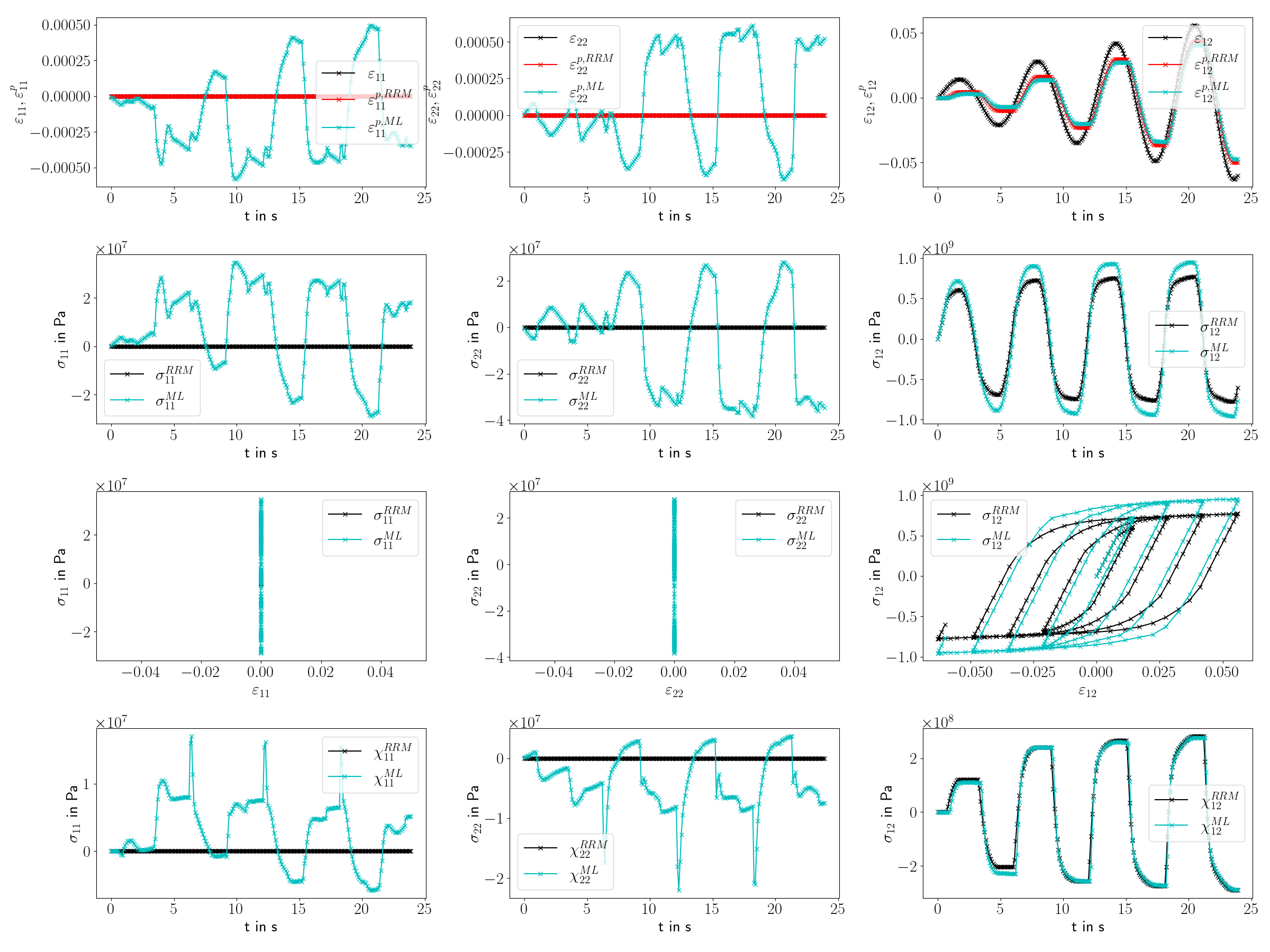}
			\subcaption{}
			\label{fig:results60102_6F}
		\end{subfigure}
		\hfill
		\begin{subfigure}{0.49\linewidth}
			\centering
			\includegraphics[trim={41.5cm 11.5cm 0.2cm 22cm},clip,width=\linewidth]{results__60102_test_No_6}
			\subcaption{}
			\label{fig:results60102_6I}
		\end{subfigure}
		\caption{Shear stress results (ML vs. RRM) for a pure shear loading test in 12 direction, model B. a) Stress-time curve. b) Stress-strain curve.}
		\label{fig:results60102_6FI}
	\end{figure}

	\paragraph{Pure shear test}
	The pure shear test considers prescribed shear strains with linearly increasing amplitude
	\begin{align}
		\varepsilon_{12}(t) = f_s \left(a + \frac{t}{t_{end}}\right) \cdot \sin(t) \quad 
	\end{align}
	with the scaling factor $f_s = 5  \, \frac{\sigma_{Y}}{E}$, the initial proportionality constant $a = 0.2$ and time step width $\Delta t = 0.1$. This test shows the highest deviations compared to all evaluated loading histories. The amplitudes of shear stresses are stably overestimated for about 19\%, independently from the artificial time step width (Fig. \ref{fig:results60102_6FI}).
	
	\paragraph{Multiaxial test}
	The following strain history is prescribed to investigate the model behabior for complex cyclic deformation
	\begin{align}
				\varepsilon_{11}(t) \quad & = \quad f_s \left(a + \frac{t}{t_{end}}\right) \cdot \sin(t)\\
				\varepsilon_{22}(t) = \varepsilon_{33}(t) \quad & = \quad   f_s \left(a + \frac{t}{t_{end}}\right) \cdot \sin(c \cdot t)\\
				\varepsilon_{12}(t) = \varepsilon_{13}(t) = \varepsilon_{23}(t) \quad & = \quad  f_s \left(a + \frac{t}{t_{end}}\right) \cdot \left( b - \cos(t) \right) \quad .\\
	\end{align}
	Here, the parameters are set as follows $f_s = 6  \, \frac{\sigma_{Y}}{E}$, $a = 0.2$, $b = 0.5$ and $c = 0.5$.
	
	The results, especially normal stress components $11$, $22$, $33$ and back stresses in all directions, exhibit very high accuracy. The only exception are shear stresses. They are permanently overestimated  with a discrepancy of about $20\%$ at peak values. This numerical error can be explained by the fact that the shear stresses are two orders of magnitude smaller than the normal stresses. The excellent numerical accuracy indicates that the model proposed is suitable for arbitrary strain paths.
	
	\begin{figure}[h]
		\centering
		\begin{subfigure}[c]{0.49\linewidth}
			\centering
			\includegraphics[trim={1.5cm 23cm 60cm 11cm},clip,width=\linewidth]{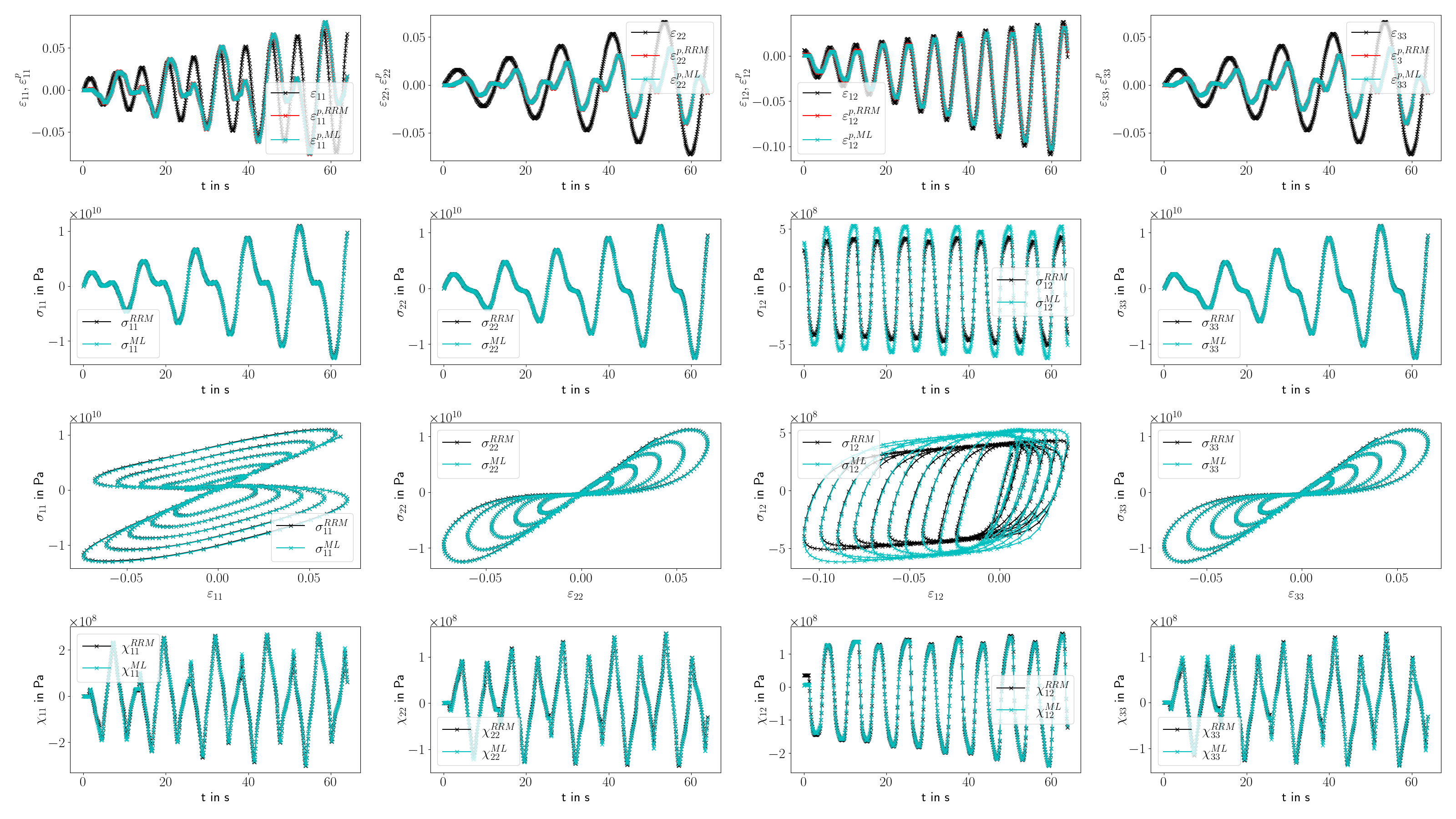}
			\subcaption{}
			\label{fig:results60102_12_sig11}
		\end{subfigure}
		\hfill
		\begin{subfigure}[c]{0.49\linewidth}
			\centering
			\includegraphics[trim={22cm 23cm 40cm 11cm},clip,width=\linewidth]{results__60102_test_No_12}
			\subcaption{}
			\label{fig:results60102_sig22}
		\end{subfigure}
		
		\begin{subfigure}[c]{0.49\linewidth}
			\centering
			\includegraphics[trim={62cm 23cm 0.2cm 11cm},clip,width=\linewidth]{results__60102_test_No_12}
			\subcaption{}
			\label{fig:results60102_12_sig33}
		\end{subfigure}
		\hfill
		\begin{subfigure}[c]{0.49\linewidth}
			\centering
			\includegraphics[trim={42cm 23cm 20cm 11cm},clip,width=\linewidth]{results__60102_test_No_12}
			\subcaption{}
			\label{fig:results60102_12_sig12}
		\end{subfigure}

		\begin{subfigure}[c]{0.49\linewidth}
			\centering
			\includegraphics[trim={1.5cm 0.2cm 60cm 33.5cm},clip,width=\linewidth]{results__60102_test_No_12}
			\subcaption{}
			\label{fig:results60102_12_chi11}
		\end{subfigure}
		\hfill
		\begin{subfigure}[c]{0.49\linewidth}
			\centering
			\includegraphics[trim={22cm 0.2cm 40cm 33.5cm},clip,width=\linewidth]{results__60102_test_No_12}
			\subcaption{}
			\label{fig:results60102_chi22}
		\end{subfigure}
		
		\begin{subfigure}[c]{0.49\linewidth}
			\centering
			\includegraphics[trim={62cm 0.2cm 0.2cm 33.5cm},clip,width=\linewidth]{results__60102_test_No_12}
			\subcaption{}
			\label{fig:results60102_12_chi33}
		\end{subfigure}
		\hfill
		\begin{subfigure}[c]{0.49\linewidth}
			\centering
			\includegraphics[trim={42cm 0.2cm 20cm 33.5cm},clip,width=\linewidth]{results__60102_test_No_12}
			\subcaption{}
			\label{fig:results60102_12_chi12}
		\end{subfigure}	
		
		\caption{
			Stress results (ML vs. RRM) for a multiaxial deformation test, model B.
		}
		\label{fig:results60102_12FI}
	\end{figure}
	
	\section{Conclusions and Outlook}\label{sec:ConcOut}
	
	The present paper investigates the application of stateless RNNs for the simulation of cyclic plasticity. It compares three models with different assumptions for the loss function. In addition to the data-driven part, regularizations include the conditions on the deviatoric characteristic of the plastic strains and back stresses, compliance with the yield criterion, the evolution of plastic strains and the associativity of the yield function. However, the analysis of the results shows that the latter has a negligible influence on the accuracy which indicates that the model is also applicable for the non-associative plasticity.
	
	Compared to existing models dealing with two-dimensional examples, the proposed NN architecture is simpler and thus more efficient. The training data sets are of smaller extent while training a complete three-dimensional material model.
	Along with the physics-informed regularization, further
	prerequisites for this superior performance are the normalization of all NN inputs and outputs, maximal possible off-loading of the NN, and the selection of a stateless network architecture with explainable feedback quantities. 
	
	In a future work, improvements on the model are expected to reduce the training effort and remaining deviations in the back and shear stress results.
	To this end, the following approaches will be tested: i) The symmetry between the coordinate axes can be used for data augmentation. ii) The redundancies of the NN outputs will be identified and exploited to shrink the NN architecture. iii) A better efficiency will be gained from controlling the sampling for data generation towards an equal coverage of the input space.
	
	On the other hand, the future work will also demonstrate the generalizability of the approach to the non-associative flow. Moreover, it will focus on the extension of the RNN based architectures that is needed for the implementation of additional physical phenomena such as temperature influences or damage evolution.
	
	\clearpage
	\backmatter
	
	\bmhead{Abbreviations}
	\begin{itemize}
		\item CANN \dots Constitutive Artificial Neural Network
		\item FEM \dots Finite Element Method
		\item GRU \dots Gated Recurrent Unit
		\item LSTM \dots Long Short-Term Memory
		\item ML \dots Machine Learning
		\item NN \dots Neural Network
		\item RNN \dots Recurrent Neural Network
		\item RRM \dots Radial Return Mapping
		\item RVE \dots Representative Volume Element
		\item FCNN \dots Fully Connected Neural Network
		
	\end{itemize}

	%
	%
	%

	\bmhead{Acknowledgments}
	
	The authors are grateful to the IT department of the Institute of Mathematics of TU Berlin for providing its facilities for HPC calculations.
	
	The authors want to express their thanks to Surya Kalidindi and his team at the Rae S. and Frank H. Neely Chair at Georgia Institute of Technology as well as Lukas Morand and Dirk Helm for the helpful discussions on the topic.
	
	\section*{Declarations}	
	\begin{itemize}
		\item Funding -- The authors received no financial support for the research, authorship, and/or publication of this article. 
		\item Conflicts of interest/Competing interests -- The authors declare that there is no conflict of interest.
		\item Ethics approval -- Not applicable
		\item Consent to participate -- Not applicable
		\item Consent for publication -- Not applicable
		\item Availability of data and material -- Data are generated numerically by using explained conventional methods.
		\item Code availability -- All used libraries are Open Source Software. The reviewed source code is made publicly available by their respective authors. Relevant changes to these codes are described in the paper. 
		\item Authors' contributions -- S. Hildebrand: Data curation, Formal Analysis, Investigation, Methodology, Software, Validation, Visualization, Writing -- original draft; S. Klinge: Conceptualization, Project administration, Resources, Supervision, Visualization, Writing -- review \& editing
	\end{itemize}

\bibliography{Paper1}	


\begin{thebibliography}{42}
\ifx \bisbn   \undefined \def \bisbn  #1{ISBN #1}\fi
\ifx \binits  \undefined \def \binits#1{#1}\fi
\ifx \bauthor  \undefined \def \bauthor#1{#1}\fi
\ifx \batitle  \undefined \def \batitle#1{#1}\fi
\ifx \bjtitle  \undefined \def \bjtitle#1{#1}\fi
\ifx \bvolume  \undefined \def \bvolume#1{\textbf{#1}}\fi
\ifx \byear  \undefined \def \byear#1{#1}\fi
\ifx \bissue  \undefined \def \bissue#1{#1}\fi
\ifx \bfpage  \undefined \def \bfpage#1{#1}\fi
\ifx \blpage  \undefined \def \blpage #1{#1}\fi
\ifx \burl  \undefined \def \burl#1{\textsf{#1}}\fi
\ifx \doiurl  \undefined \def \doiurl#1{\url{https://doi.org/#1}}\fi
\ifx \betal  \undefined \def \betal{\textit{et al.}}\fi
\ifx \binstitute  \undefined \def \binstitute#1{#1}\fi
\ifx \binstitutionaled  \undefined \def \binstitutionaled#1{#1}\fi
\ifx \bctitle  \undefined \def \bctitle#1{#1}\fi
\ifx \beditor  \undefined \def \beditor#1{#1}\fi
\ifx \bpublisher  \undefined \def \bpublisher#1{#1}\fi
\ifx \bbtitle  \undefined \def \bbtitle#1{#1}\fi
\ifx \bedition  \undefined \def \bedition#1{#1}\fi
\ifx \bseriesno  \undefined \def \bseriesno#1{#1}\fi
\ifx \blocation  \undefined \def \blocation#1{#1}\fi
\ifx \bsertitle  \undefined \def \bsertitle#1{#1}\fi
\ifx \bsnm \undefined \def \bsnm#1{#1}\fi
\ifx \bsuffix \undefined \def \bsuffix#1{#1}\fi
\ifx \bparticle \undefined \def \bparticle#1{#1}\fi
\ifx \barticle \undefined \def \barticle#1{#1}\fi
\bibcommenthead
\ifx \bconfdate \undefined \def \bconfdate #1{#1}\fi
\ifx \botherref \undefined \def \botherref #1{#1}\fi
\ifx \url \undefined \def \url#1{\textsf{#1}}\fi
\ifx \bchapter \undefined \def \bchapter#1{#1}\fi
\ifx \bbook \undefined \def \bbook#1{#1}\fi
\ifx \bcomment \undefined \def \bcomment#1{#1}\fi
\ifx \oauthor \undefined \def \oauthor#1{#1}\fi
\ifx \citeauthoryear \undefined \def \citeauthoryear#1{#1}\fi
\ifx \endbibitem  \undefined \def \endbibitem {}\fi
\ifx \bconflocation  \undefined \def \bconflocation#1{#1}\fi
\ifx \arxivurl  \undefined \def \arxivurl#1{\textsf{#1}}\fi
\csname PreBibitemsHook\endcsname

\bibitem[\protect\citeauthoryear{Rosenkranz et~al.}{2023}]{Rosenkranz2023}
\begin{botherref}
\oauthor{\bsnm{Rosenkranz}, \binits{M.}},
\oauthor{\bsnm{Kalina}, \binits{K.A.}},
\oauthor{\bsnm{Brummund}, \binits{J.}},
\oauthor{\bsnm{Kästner}, \binits{M.}}:
A comparative study on different neural network architectures to model
  inelasticity.
International Journal for Numerical Methods in Engineering
(124),
4802--4840
(2023)
\doiurl{10.1002/nme.7319}
{\href{https://arxiv.org/abs/https://onlinelibrary.wiley.com/doi/pdf/10.1002/nme.7319}{{https://onlinelibrary.wiley.com/doi/pdf/10.1002/nme.7319}}}
\end{botherref}
\endbibitem

\bibitem[\protect\citeauthoryear{Dornheim et~al.}{2023}]{dornheim2023neural}
\begin{botherref}
\oauthor{\bsnm{Dornheim}, \binits{J.}},
\oauthor{\bsnm{Morand}, \binits{L.}},
\oauthor{\bsnm{Nallani}, \binits{H.J.}},
\oauthor{\bsnm{Helm}, \binits{D.}}:
Neural Networks for Constitutive Modeling -- From Universal Function
  Approximators to Advanced Models and the Integration of Physics
(2023)
\end{botherref}
\endbibitem

\bibitem[\protect\citeauthoryear{Bock et~al.}{2019}]{Bock2019}
\begin{botherref}
\oauthor{\bsnm{Bock}, \binits{F.E.}},
\oauthor{\bsnm{Aydin}, \binits{R.C.}},
\oauthor{\bsnm{Cyron}, \binits{C.J.}},
\oauthor{\bsnm{Huber}, \binits{N.}},
\oauthor{\bsnm{Kalidindi}, \binits{S.R.}},
\oauthor{\bsnm{Klusemann}, \binits{B.}}:
A review of the application of machine learning and data mining approaches in
  continuum materials mechanics.
Frontiers in Materials
\textbf{6}
(2019)
\doiurl{10.3389/fmats.2019.00110}
\end{botherref}
\endbibitem

\bibitem[\protect\citeauthoryear{Lei et~al.}{2024}]{Lei2024}
\begin{barticle}
\bauthor{\bsnm{Lei}, \binits{M.}},
\bauthor{\bsnm{Sun}, \binits{G.}},
\bauthor{\bsnm{Yang}, \binits{G.}},
\bauthor{\bsnm{Wen}, \binits{B.}}:
\batitle{A computational mechanical constitutive modeling method based on
  thermally-activated microstructural evolution and strengthening mechanisms}.
\bjtitle{International Journal of Plasticity}
\bvolume{173},
\bfpage{103881}
(\byear{2024})
\doiurl{10.1016/j.ijplas.2024.103881}
\end{barticle}
\endbibitem

\bibitem[\protect\citeauthoryear{Miehe et~al.}{2010}]{Miehe2010}
\begin{barticle}
\bauthor{\bsnm{Miehe}, \binits{C.}},
\bauthor{\bsnm{Hofacker}, \binits{M.}},
\bauthor{\bsnm{Welschinger}, \binits{F.}}:
\batitle{A phase field model for rate-independent crack propagation: Robust
  algorithmic implementation based on operator splits}.
\bjtitle{Computer Methods in Applied Mechanics and Engineering}
\bvolume{199}(\bissue{45}),
\bfpage{2765}--\blpage{2778}
(\byear{2010})
\doiurl{10.1016/j.cma.2010.04.011}
\end{barticle}
\endbibitem

\bibitem[\protect\citeauthoryear{Aydiner et~al.}{2024}]{Aydiner2024}
\begin{barticle}
\bauthor{\bsnm{Aydiner}, \binits{I.U.}},
\bauthor{\bsnm{Tatli}, \binits{B.}},
\bauthor{\bsnm{Yalçinkaya}, \binits{T.}}:
\batitle{Investigation of failure mechanisms in dual-phase steels through
  cohesive zone modeling and crystal plasticity frameworks}.
\bjtitle{International Journal of Plasticity}
\bvolume{174},
\bfpage{103898}
(\byear{2024})
\doiurl{10.1016/j.ijplas.2024.103898}
\end{barticle}
\endbibitem

\bibitem[\protect\citeauthoryear{Bartošák and Horváth}{2024}]{Bartosak2024}
\begin{barticle}
\bauthor{\bsnm{Bartošák}, \binits{M.}},
\bauthor{\bsnm{Horváth}, \binits{J.}}:
\batitle{A continuum damage coupled unified viscoplastic model for simulating
  the mechanical behaviour of a ductile cast iron under isothermal low-cycle
  fatigue, fatigue-creep and creep loading}.
\bjtitle{International Journal of Plasticity}
\bvolume{173},
\bfpage{103868}
(\byear{2024})
\doiurl{10.1016/j.ijplas.2023.103868}
\end{barticle}
\endbibitem

\bibitem[\protect\citeauthoryear{Hornik et~al.}{1989}]{Hornik1989}
\begin{barticle}
\bauthor{\bsnm{Hornik}, \binits{K.}},
\bauthor{\bsnm{Stinchcombe}, \binits{M.}},
\bauthor{\bsnm{White}, \binits{H.}}:
\batitle{Multilayer feedforward networks are universal approximators}.
\bjtitle{Neural Networks}
\bvolume{2}(\bissue{5}),
\bfpage{359}--\blpage{366}
(\byear{1989})
\doiurl{10.1016/0893-6080(89)90020-8}
\end{barticle}
\endbibitem

\bibitem[\protect\citeauthoryear{Raissi et~al.}{2019}]{Raissi2019}
\begin{barticle}
\bauthor{\bsnm{Raissi}, \binits{M.}},
\bauthor{\bsnm{Perdikaris}, \binits{P.}},
\bauthor{\bsnm{Karniadakis}, \binits{G.E.}}:
\batitle{Physics-informed neural networks: A deep learning framework for
  solving forward and inverse problems involving nonlinear partial differential
  equations}.
\bjtitle{Journal of Computational Physics}
\bvolume{378},
\bfpage{686}--\blpage{707}
(\byear{2019})
\doiurl{10.1016/j.jcp.2018.10.045}
\end{barticle}
\endbibitem

\bibitem[\protect\citeauthoryear{Kollmannsberger}{2021}]{Kollmannsberger2021}
\begin{bbook}
\bauthor{\bsnm{Kollmannsberger}, \binits{S.}}:
\bbtitle{Deep Learning in Computational Mechanics: an Introductory Course}.
\bsertitle{Studies in computational intelligence}.
\bpublisher{Springer},
\blocation{Cham}
(\byear{2021}).
\burl{https://link.springer.com/10.1007/978-3-030-76587-3}
\end{bbook}
\endbibitem

\bibitem[\protect\citeauthoryear{Hildebrand and
  Klinge}{2023}]{hildebrand2023comparison}
\begin{botherref}
\oauthor{\bsnm{Hildebrand}, \binits{S.}},
\oauthor{\bsnm{Klinge}, \binits{S.}}:
Comparison of Neural FEM and Neural Operator Methods for applications in Solid
  Mechanics
(2023)
\end{botherref}
\endbibitem

\bibitem[\protect\citeauthoryear{Bock et~al.}{2021}]{BockKeller2021}
\begin{botherref}
\oauthor{\bsnm{Bock}, \binits{F.E.}},
\oauthor{\bsnm{Keller}, \binits{S.}},
\oauthor{\bsnm{Huber}, \binits{N.}},
\oauthor{\bsnm{Klusemann}, \binits{B.}}:
Hybrid modelling by machine learning corrections of analytical model
  predictions towards high-fidelity simulation solutions.
Materials
\textbf{14}(8)
(2021)
\doiurl{10.3390/ma14081883}
\end{botherref}
\endbibitem

\bibitem[\protect\citeauthoryear{Simo and Hughes}{1998}]{SimoHughes}
\begin{bbook}
\bauthor{\bsnm{Simo}, \binits{J.C.}},
\bauthor{\bsnm{Hughes}, \binits{T.J.R.}}:
\bbtitle{Computational Inelasticity}.
\bsertitle{Interdisciplinary Applied Mathematics},
vol. \bseriesno{7}.
\bpublisher{Springer},
\blocation{New York, Berlin, Heidelberg}
(\byear{1998})
\end{bbook}
\endbibitem

\bibitem[\protect\citeauthoryear{Khan and Huang}{1995}]{KhanHuang}
\begin{bbook}
\bauthor{\bsnm{Khan}, \binits{A.S.}},
\bauthor{\bsnm{Huang}, \binits{S.}}:
\bbtitle{Continuum Theory of Plasticity}.
\bpublisher{John Wiley and Sons, Inc.},
\blocation{New York, Chichester, Brisbane, Toronto, Singapore}
(\byear{1995})
\end{bbook}
\endbibitem

\bibitem[\protect\citeauthoryear{Bland}{1957}]{Bland1957}
\begin{barticle}
\bauthor{\bsnm{Bland}, \binits{D.R.}}:
\batitle{The associated flow rule of plasticity}.
\bjtitle{Journal of the Mechanics and Physics of Solids}
\bvolume{6}(\bissue{1}),
\bfpage{71}--\blpage{78}
(\byear{1957})
\doiurl{10.1016/0022-5096(57)90049-2}
\end{barticle}
\endbibitem

\bibitem[\protect\citeauthoryear{Suchocki}{2022}]{Suchocki2021}
\begin{barticle}
\bauthor{\bsnm{Suchocki}, \binits{C.}}:
\batitle{On finite element implementation of cyclic elastoplasticity: theory,
  coding, and exemplary problems}.
\bjtitle{Acta Mechanica}
\bvolume{233},
\bfpage{83}--\blpage{120}
(\byear{2022})
\doiurl{10.1007/s00707-021-03069-3}
\end{barticle}
\endbibitem

\bibitem[\protect\citeauthoryear{Frederick and
  Armstrong}{2007}]{Frederick-Armstrong2007}
\begin{barticle}
\bauthor{\bsnm{Frederick}, \binits{C.O.}},
\bauthor{\bsnm{Armstrong}, \binits{P.J.}}:
\batitle{A mathematical representation of the multiaxial bauschinger effect}.
\bjtitle{Materials at High Temperatures}
\bvolume{24},
\bfpage{1}--\blpage{26}
(\byear{2007})
\end{barticle}
\endbibitem

\bibitem[\protect\citeauthoryear{Aygün et~al.}{2021}]{AyguenKlinge2021}
\begin{barticle}
\bauthor{\bsnm{Aygün}, \binits{S.}},
\bauthor{\bsnm{Wiegold}, \binits{T.}},
\bauthor{\bsnm{Klinge}, \binits{S.}}:
\batitle{Coupling of the phase field approach to the armstrong-frederick model
  for the simulation of ductile damage under cyclic load}.
\bjtitle{International Journal of Plasticity}
\bvolume{143},
\bfpage{103021}
(\byear{2021})
\doiurl{10.1016/j.ijplas.2021.103021}
\end{barticle}
\endbibitem

\bibitem[\protect\citeauthoryear{Furukawa and
  Hoffman}{2004}]{FurukawaHoffman2004}
\begin{barticle}
\bauthor{\bsnm{Furukawa}, \binits{T.}},
\bauthor{\bsnm{Hoffman}, \binits{M.}}:
\batitle{Accurate cyclic plastic analysis using a neural network material
  model}.
\bjtitle{Engineering Analysis with Boundary Elements}
\bvolume{28}(\bissue{3}),
\bfpage{195}--\blpage{204}
(\byear{2004})
\doiurl{10.1016/S0955-7997(03)00050-X}
\end{barticle}
\endbibitem

\bibitem[\protect\citeauthoryear{Gorji et~al.}{2020}]{MohrGorji2020}
\begin{barticle}
\bauthor{\bsnm{Gorji}, \binits{M.B.}},
\bauthor{\bsnm{Mozaffar}, \binits{M.}},
\bauthor{\bsnm{Heidenreich}, \binits{J.N.}},
\bauthor{\bsnm{Cao}, \binits{J.}},
\bauthor{\bsnm{Mohr}, \binits{D.}}:
\batitle{On the potential of recurrent neural networks for modeling path
  dependent plasticity}.
\bjtitle{Journal of the Mechanics and Physics of Solids}
\bvolume{143},
\bfpage{103972}
(\byear{2020})
\doiurl{10.1016/j.jmps.2020.103972}
\end{barticle}
\endbibitem

\bibitem[\protect\citeauthoryear{Jang et~al.}{2021}]{Jang2021}
\begin{barticle}
\bauthor{\bsnm{Jang}, \binits{D.P.}},
\bauthor{\bsnm{Fazily}, \binits{P.}},
\bauthor{\bsnm{Yoon}, \binits{J.W.}}:
\batitle{Machine learning-based constitutive model for j2- plasticity}.
\bjtitle{International Journal of Plasticity}
\bvolume{138},
\bfpage{102919}
(\byear{2021})
\doiurl{10.1016/j.ijplas.2020.102919}
\end{barticle}
\endbibitem

\bibitem[\protect\citeauthoryear{Huang et~al.}{2020}]{HuangWriggers2020}
\begin{barticle}
\bauthor{\bsnm{Huang}, \binits{D.}},
\bauthor{\bsnm{Fuhg}, \binits{J.N.}},
\bauthor{\bsnm{Weißenfels}, \binits{C.}},
\bauthor{\bsnm{Wriggers}, \binits{P.}}:
\batitle{A machine learning based plasticity model using proper orthogonal
  decomposition}.
\bjtitle{Computer Methods in Applied Mechanics and Engineering}
\bvolume{365},
\bfpage{113008}
(\byear{2020})
\doiurl{10.1016/j.cma.2020.113008}
\end{barticle}
\endbibitem

\bibitem[\protect\citeauthoryear{Logarzo et~al.}{2021}]{Rimoli2021}
\begin{barticle}
\bauthor{\bsnm{Logarzo}, \binits{H.J.}},
\bauthor{\bsnm{Capuano}, \binits{G.}},
\bauthor{\bsnm{Rimoli}, \binits{J.J.}}:
\batitle{Smart constitutive laws: Inelastic homogenization through machine
  learning}.
\bjtitle{Computer Methods in Applied Mechanics and Engineering}
\bvolume{373},
\bfpage{113482}
(\byear{2021})
\doiurl{10.1016/j.cma.2020.113482}
\end{barticle}
\endbibitem

\bibitem[\protect\citeauthoryear{Cho et~al.}{2014}]{GRU2014}
\begin{botherref}
\oauthor{\bsnm{Cho}, \binits{K.}},
\oauthor{\bsnm{Merrienboer}, \binits{B.}},
\oauthor{\bsnm{Bahdanau}, \binits{D.}},
\oauthor{\bsnm{Bengio}, \binits{Y.}}:
On the properties of neural machine translation: Encoder-decoder approaches.
CoRR
\textbf{abs/1409.1259}
(2014)
{\href{https://arxiv.org/abs/1409.1259}{{1409.1259}}}
\end{botherref}
\endbibitem

\bibitem[\protect\citeauthoryear{Bonatti and Mohr}{2021}]{BonattiMohr2021}
\begin{barticle}
\bauthor{\bsnm{Bonatti}, \binits{C.}},
\bauthor{\bsnm{Mohr}, \binits{D.}}:
\batitle{One for all: Universal material model based on minimal state-space
  neural networks}.
\bjtitle{Science Advances}
\bvolume{7}(\bissue{26}),
\bfpage{3658}
(\byear{2021})
\doiurl{10.1126/sciadv.abf3658}
{\href{https://arxiv.org/abs/https://www.science.org/doi/pdf/10.1126/sciadv.abf3658}{{https://www.science.org/doi/pdf/10.1126/sciadv.abf3658}}}
\end{barticle}
\endbibitem

\bibitem[\protect\citeauthoryear{{Joudivand Sarand} and
  Misirlioglu}{2024}]{JoudivandSarand2024}
\begin{barticle}
\bauthor{\bsnm{{Joudivand Sarand}}, \binits{M.H.}},
\bauthor{\bsnm{Misirlioglu}, \binits{I.B.}}:
\batitle{A physics-based plasticity study of the mechanism of inhomogeneous
  strain evolution in dual phase 600 steel}.
\bjtitle{International Journal of Plasticity}
\bvolume{174},
\bfpage{103918}
(\byear{2024})
\doiurl{10.1016/j.ijplas.2024.103918}
\end{barticle}
\endbibitem

\bibitem[\protect\citeauthoryear{Linka et~al.}{2021}]{LinkaCyron2021}
\begin{barticle}
\bauthor{\bsnm{Linka}, \binits{K.}},
\bauthor{\bsnm{Hillgärtner}, \binits{M.}},
\bauthor{\bsnm{Abdolazizi}, \binits{K.P.}},
\bauthor{\bsnm{Aydin}, \binits{R.C.}},
\bauthor{\bsnm{Itskov}, \binits{M.}},
\bauthor{\bsnm{Cyron}, \binits{C.J.}}:
\batitle{Constitutive artificial neural networks: A fast and general approach
  to predictive data-driven constitutive modeling by deep learning}.
\bjtitle{Journal of Computational Physics}
\bvolume{429},
\bfpage{110010}
(\byear{2021})
\doiurl{10.1016/j.jcp.2020.110010}
\end{barticle}
\endbibitem

\bibitem[\protect\citeauthoryear{As'ad et~al.}{2022}]{Farhat2022}
\begin{barticle}
\bauthor{\bsnm{As'ad}, \binits{F.}},
\bauthor{\bsnm{Avery}, \binits{P.}},
\bauthor{\bsnm{Farhat}, \binits{C.}}:
\batitle{A mechanics-informed artificial neural network approach in data-driven
  constitutive modeling}.
\bjtitle{International Journal for Numerical Methods in Engineering}
\bvolume{123}(\bissue{12}),
\bfpage{2738}--\blpage{2759}
(\byear{2022})
\doiurl{10.1002/nme.6957}
{\href{https://arxiv.org/abs/https://onlinelibrary.wiley.com/doi/pdf/10.1002/nme.6957}{{https://onlinelibrary.wiley.com/doi/pdf/10.1002/nme.6957}}}
\end{barticle}
\endbibitem

\bibitem[\protect\citeauthoryear{Linka and Kuhl}{2023}]{LinkaKuhl2023}
\begin{barticle}
\bauthor{\bsnm{Linka}, \binits{K.}},
\bauthor{\bsnm{Kuhl}, \binits{E.}}:
\batitle{A new family of constitutive artificial neural networks towards
  automated model discovery}.
\bjtitle{Computer Methods in Applied Mechanics and Engineering}
\bvolume{403},
\bfpage{115731}
(\byear{2023})
\doiurl{10.1016/j.cma.2022.115731}
\end{barticle}
\endbibitem

\bibitem[\protect\citeauthoryear{Zhang and Mohr}{2020}]{ZhangMohr2020}
\begin{barticle}
\bauthor{\bsnm{Zhang}, \binits{A.}},
\bauthor{\bsnm{Mohr}, \binits{D.}}:
\batitle{Using neural networks to represent von mises plasticity with isotropic
  hardening}.
\bjtitle{International Journal of Plasticity}
\bvolume{132},
\bfpage{102732}
(\byear{2020})
\doiurl{10.1016/j.ijplas.2020.102732}
\end{barticle}
\endbibitem

\bibitem[\protect\citeauthoryear{Shoghi and
  Hartmaier}{2022}]{ShoghiHartmaier2022}
\begin{botherref}
\oauthor{\bsnm{Shoghi}, \binits{R.}},
\oauthor{\bsnm{Hartmaier}, \binits{A.}}:
Optimal data-generation strategy for machine learning yield functions in
  anisotropic plasticity.
Frontiers in Materials
\textbf{9}
(2022)
\doiurl{10.3389/fmats.2022.868248}
\end{botherref}
\endbibitem

\bibitem[\protect\citeauthoryear{Shoghi et~al.}{2024}]{Shoghi2024}
\begin{botherref}
\oauthor{\bsnm{Shoghi}, \binits{R.}},
\oauthor{\bsnm{Morand}, \binits{L.}},
\oauthor{\bsnm{Helm}, \binits{D.}},
\oauthor{\bsnm{Hartmaier}, \binits{A.}}:
Optimizing machine learning yield functions using query-by-committee for
  support vector classification with a dynamic stopping criterion.
Computational Mechanics
(2024)
\end{botherref}
\endbibitem

\bibitem[\protect\citeauthoryear{Lubliner et~al.}{1989}]{Lubliner1989}
\begin{barticle}
\bauthor{\bsnm{Lubliner}, \binits{J.}},
\bauthor{\bsnm{Oliver}, \binits{J.}},
\bauthor{\bsnm{Oller}, \binits{S.}},
\bauthor{\bsnm{Oñate}, \binits{E.}}:
\batitle{A plastic-damage model for concrete}.
\bjtitle{International Journal of Solids and Structures}
\bvolume{25}(\bissue{3}),
\bfpage{299}--\blpage{326}
(\byear{1989})
\doiurl{10.1016/0020-7683(89)90050-4}
\end{barticle}
\endbibitem

\bibitem[\protect\citeauthoryear{Vacev et~al.}{2023}]{Zoric2023}
\begin{barticle}
\bauthor{\bsnm{Vacev}, \binits{T.}},
\bauthor{\bsnm{Zorić}, \binits{A.}},
\bauthor{\bsnm{Grdić}, \binits{D.}},
\bauthor{\bsnm{Ristić}, \binits{N.}},
\bauthor{\bsnm{Grdić}, \binits{Z.}},
\bauthor{\bsnm{Milić}, \binits{M.}}:
\batitle{Experimental and numerical analysis of impact strength of concrete
  slabs}.
\bjtitle{Periodica Polytechnica Civil Engineering}
\bvolume{67}(\bissue{1}),
\bfpage{325}--\blpage{335}
(\byear{2023})
\doiurl{10.3311/PPci.21084}
\end{barticle}
\endbibitem

\bibitem[\protect\citeauthoryear{Oliveira and Penna}{2004}]{Oliveira2022}
\begin{barticle}
\bauthor{\bsnm{Oliveira}, \binits{D.B.}},
\bauthor{\bsnm{Penna}, \binits{S.S.}}:
\batitle{A general framework for finite strain elastoplastic models: a
  theoretical approach}.
\bjtitle{Journal of the Brazilian Society of Mechanical Sciences and
  Engineering}
\bvolume{44},
\bfpage{87}--\blpage{116}
(\byear{2004})
\doiurl{10.1007/s40430-022-03647-z}
\end{barticle}
\endbibitem

\bibitem[\protect\citeauthoryear{Dettmer and Reese}{2004}]{Reese2004}
\begin{barticle}
\bauthor{\bsnm{Dettmer}, \binits{W.}},
\bauthor{\bsnm{Reese}, \binits{S.}}:
\batitle{On the theoretical and numerical modelling of armstrong–frederick
  kinematic hardening in the finite strain regime}.
\bjtitle{Computer Methods in Applied Mechanics and Engineering}
\bvolume{193}(\bissue{1}),
\bfpage{87}--\blpage{116}
(\byear{2004})
\doiurl{10.1016/j.cma.2003.09.005}
\end{barticle}
\endbibitem

\bibitem[\protect\citeauthoryear{Kingma and Ba}{2014}]{Kingma2014}
\begin{botherref}
\oauthor{\bsnm{Kingma}, \binits{D.P.}},
\oauthor{\bsnm{Ba}, \binits{J.}}:
Adam: A Method for Stochastic Optimization.
arXiv
(2014).
\doiurl{10.48550/ARXIV.1412.6980} .
\url{https://arxiv.org/abs/1412.6980}
\end{botherref}
\endbibitem

\bibitem[\protect\citeauthoryear{Paszke et~al.}{2017}]{Paszke2017}
\begin{bchapter}
\bauthor{\bsnm{Paszke}, \binits{A.}},
\bauthor{\bsnm{Gross}, \binits{S.}},
\bauthor{\bsnm{Chintala}, \binits{S.}},
\bauthor{\bsnm{Chanan}, \binits{G.}},
\bauthor{\bsnm{Yang}, \binits{E.}},
\bauthor{\bsnm{DeVito}, \binits{Z.}},
\bauthor{\bsnm{Lin}, \binits{Z.}},
\bauthor{\bsnm{Desmaison}, \binits{A.}},
\bauthor{\bsnm{Antiga}, \binits{L.}},
\bauthor{\bsnm{Lerer}, \binits{A.}}:
\bctitle{Automatic differentiation in pytorch}.
In: \bbtitle{Autodiff Workshop NIPS 2017}
(\byear{2017}).
\burl{https://api.semanticscholar.org/CorpusID:40027675}
\end{bchapter}
\endbibitem

\bibitem[\protect\citeauthoryear{Nguyen-Thanh et~al.}{2020}]{Nguyen-Thanh2020}
\begin{barticle}
\bauthor{\bsnm{Nguyen-Thanh}, \binits{V.M.}},
\bauthor{\bsnm{Zhuang}, \binits{X.}},
\bauthor{\bsnm{Rabczuk}, \binits{T.}}:
\batitle{A deep energy method for finite deformation hyperelasticity}.
\bjtitle{European Journal of Mechanics - A/Solids}
\bvolume{80},
\bfpage{103874}
(\byear{2020})
\doiurl{10.1016/j.euromechsol.2019.103874}
\end{barticle}
\endbibitem

\bibitem[\protect\citeauthoryear{Krishnapriyan et~al.}{}]{Krishnapriyan2021}
\begin{botherref}
\oauthor{\bsnm{Krishnapriyan}, \binits{A.}},
\oauthor{\bsnm{Gholami}, \binits{A.}},
\oauthor{\bsnm{Zhe}, \binits{S.}},
\oauthor{\bsnm{Kirby}, \binits{R.}},
\oauthor{\bsnm{Mahoney}, \binits{M.W.}}:
Characterizing possible failure modes in physics-informed neural networks.
In: Ranzato, M., Beygelzimer, A., Dauphin, Y., Liang, P.S., Vaughan, J.W.
  (eds.)
Advances in Neural Information Processing Systems,
vol. 34,
pp. 26548--26560.
Curran Associates, Inc.
\url{https://proceedings.neurips.cc/paper/2021/file/df438e5206f31600e6ae4af72f2725f1-Paper.pdf}
\end{botherref}
\endbibitem

\bibitem[\protect\citeauthoryear{Wang et~al.}{2021b}]{Wang2021b}
\begin{botherref}
\oauthor{\bsnm{Wang}, \binits{S.}},
\oauthor{\bsnm{Wang}, \binits{H.}},
\oauthor{\bsnm{Perdikaris}, \binits{P.}}:
Learning the solution operator of parametric partial differential equations
  with physics-informed DeepOnets.
arXiv
(2021).
\doiurl{10.48550/ARXIV.2103.10974} .
\url{https://arxiv.org/abs/2103.10974}
\end{botherref}
\endbibitem

\bibitem[\protect\citeauthoryear{Fuhg and Bouklas}{2022}]{FuhgBouklas2022}
\begin{barticle}
\bauthor{\bsnm{Fuhg}, \binits{J.N.}},
\bauthor{\bsnm{Bouklas}, \binits{N.}}:
\batitle{The mixed deep energy method for resolving concentration features in
  finite strain hyperelasticity}.
\bjtitle{Journal of Computational Physics}
\bvolume{451},
\bfpage{110839}
(\byear{2022})
\doiurl{10.1016/j.jcp.2021.110839}
\end{barticle}
\endbibitem

\end{thebibliography}
	
\end{document}